\documentclass[9pt,arxiv,secnum]{lapreprint}

\usepackage[utf8]{inputenc}

\usepackage[capitalise,nameinlink]{cleveref}
\usepackage{siunitx}         
\usepackage{pdflscape}       
\usepackage{rotating}        
\usepackage{textgreek}       
\usepackage{gensymb}         
\usepackage[misc]{ifsym}     
\usepackage{orcidlink}       
\usepackage{listings}        
\usepackage{colortbl}        
\usepackage{tabularx}        
\usepackage{longtable}       
\usepackage{subcaption}
\usepackage{multirow}
\usepackage{snotez}          
\usepackage{csquotes}        
\usepackage{amsmath}
\usepackage{todonotes}       
\usepackage{xspace}
\usepackage[normalem]{ulem}  
\usepackage{ccaption}

\usepackage[                
    backend=biber,      
	bibstyle=apa,
	citestyle=authoryear,
	uniquelist=false,
	uniquename=false,
    natbib=true,        
    hyperref=true,      
    alldates=year,      
    isbn=false,
    eprint=false,
	doi=true,
    maxcitenames=2
]{biblatex}

\AtEveryBibitem{
    \clearfield{urlyear}
    \clearfield{urlmonth}
    \clearlist{language}
}

\newcommand{\scriptfile}{\textit{psd\_by\_trial\_type.py}\xspace}
\newcommand{\plotfile}{\textit{R2G\_PSD\_all\_subjects.png}\xspace}
\newcommand{\provfile}{\textit{R2G\_PSD\_all\_subjects.ttl}\xspace}

\newcommand{\python}{\textit{Python}\xspace}
\newcommand{\neo}{\textit{Neo}\xspace}
\newcommand{\elephant}{\textit{Elephant}\xspace}
\newcommand{\quantities}{\textit{quantities}\xspace}
\newcommand{\nix}{\textit{NIX}\xspace}
\newcommand{\joblib}{\textit{joblib}\xspace}
\newcommand{\hashlib}{\textit{hashlib}\xspace}
\newcommand{\networkx}{\textit{NetworkX}\xspace}
\newcommand{\rdflib}{\textit{RDFLib}\xspace}
\newcommand{\matplotlib}{\textit{matplotlib}\xspace}
\newcommand{\scipy}{\textit{SciPy}\xspace}
\newcommand{\numpy}{\textit{NumPy}\xspace}
\newcommand{\gephi}{\textit{Gephi}\xspace}
\newcommand{\quantity}{\textit{Quantity}\xspace}

\newcommand{\git}{\textit{git}\xspace}

\newcommand{\vistrails}{\textit{VisTrails}\xspace}
\newcommand{\noworkflow}{\textit{noWorkflow}\xspace}
\newcommand{\aiida}{\textit{AiiDA}\xspace}
\newcommand{\loni}{\textit{LONI Pipeline}\xspace}
\newcommand{\snakemake}{\textit{Snakemake}\xspace}
\newcommand{\avocado}{\textit{AVOCADO}\xspace}
\newcommand{\jupyter}{\textit{Jupyter}\xspace}
\newcommand{\caesar}{\textit{CAESAR}\xspace}
\newcommand{\ProvBook}{\textit{ProvBook}\xspace}
\newcommand{\sumatra}{\textit{Sumatra}\xspace}
\newcommand{\fairworkflows}{\textit{fairworkflows}\xspace}
\newcommand{\bennch}{\textit{beNNch}\xspace}

\newcommand{\neoSegment}{\textit{Segment}\xspace}
\newcommand{\neoBlock}{\textit{Block}\xspace}
\newcommand{\neoAnalogSignal}{\textit{AnalogSignal}\xspace}
\newcommand{\neoEvent}{\textit{Event}\xspace}
\newcommand{\neoEpoch}{\textit{Epoch}\xspace}

\newcommand{\elephantWelch}{\textit{welch\_psd}\xspace}
\newcommand{\elephantButter}{\textit{butter}\xspace}

\newcommand{\matplotlibAxes}{\textit{AxesSubplot}\xspace}
\newcommand{\matplotlibFigure}{\textit{Figure}\xspace}

\newcommand{\alpacaActivate}{\textbf{activate}\xspace}
\newcommand{\alpacaSaveProvenance}{\textbf{save\_provenance}\xspace}
\newcommand{\alpacaSettings}{\textbf{alpaca\_settings}\xspace}
\newcommand{\alpacaSessionID}{\textbf{\textit{session ID}}\xspace}
\newcommand{\alpacaExecutionID}{\textbf{\textit{execution ID}}\xspace}
\newcommand{\alpacaProvenance}{\textbf{Provenance}\xspace}
\newcommand{\alpacaProvenanceGraph}{\textbf{ProvenanceGraph}\xspace}
\newcommand{\alpacaAlpacaProvDocument}{\textbf{AlpacaProvDocument}\xspace}
\newcommand{\alpacaDataObjectEntity}{\textbf{DataObjectEntity}\xspace}
\newcommand{\alpacaFileEntity}{\textbf{FileEntity}\xspace}
\newcommand{\alpacaFunctionExecution}{\textbf{FunctionExecution}\xspace}
\newcommand{\alpacaScriptAgent}{\textbf{ScriptAgent}\xspace}

\newcommand{\alpacaHasAttribute}{\textbf{hasAttribute}\xspace}
\newcommand{\alpacaHasParameter}{\textbf{hasParameter}\xspace}
\newcommand{\alpacaNameValuePair}{\textbf{NameValuePair}\xspace}

\newcommand{\scriptPlotLfpPsd}{\textit{plot\_lfp\_psd}\xspace}

\newcommand{\hash}{\textit{hash}\xspace}

\newcommand{\videourl}{https://purl.org/alpaca/video}

\title{Facilitating the sharing of electrophysiology data analysis results through in-depth provenance capture}

\author[ \orcidlink{0000-0003-0503-5264} 1,2 \Letter]{Cristiano A. Köhler}
\author[ \orcidlink{0000-0003-3489-0542} 1]{Danylo Ulianych}
\author[ \orcidlink{0000-0003-2829-2220} 1,2]{Sonja Grün}
\author[ \orcidlink{0000-0001-6324-7164} 3,4]{Stefan Decker}
\author[ \orcidlink{0000-0003-1255-7300} 1]{Michael Denker}

\affil[1]{Institute of Neuroscience and Medicine (INM-6) and Institute for Advanced Simulation (IAS-6) and JARA Institute Brain Structure-Function Relationships (INM-10), Jülich Research Centre, Jülich, Germany}
\affil[2]{Theoretical Systems Neurobiology, RWTH Aachen University, Aachen, Germany}
\affil[3]{Chair of Databases and Information Systems, RWTH Aachen University, Aachen, Germany}
\affil[4]{Fraunhofer Institute for Applied Information Technology (FIT), Sankt Augustin, Germany}

\newcommand{\printfunding}{This work was performed as part of the Helmholtz School for Data Science in Life, Earth and Energy (HDS-LEE) and received funding from the Helmholtz Association of German Research Centres. This project has received funding from the European Union’s Horizon 2020 Framework Programme for Research and Innovation under Specific Grant Agreements No. 785907 (Human Brain Project SGA2) and 945539 (Human Brain Project SGA3), the Ministry of Culture and Science of the State of North Rhine-Westphalia, Germany (NRW-network "iBehave", grant number: NW21-049), the Joint Lab "Supercomputing and Modeling for the Human Brain", and by the Helmholtz Association Initiative and Networking Fund under project number ZT-I-0003.}

\metadata[]{\Letter\hspace{.5ex} For correspondence}{\href{mailto:c.koehler@fz-juelich.de}{c.koehler@fz-juelich.de}}

\metadata[]{Keywords}{provenance, data analysis, electrophysiology, Python, software, FAIR, workflow}

\metadata[]{Funding}{\printfunding}

\metadata[]{Competing interests}{The authors declare no competing interests.}

\leadauthor{Köhler}
\shorttitle{Facilitating sharing analysis results through in-depth provenance}

\begin{document}

\maketitle

%
%

\begin{abstract}
Scientific research demands reproducibility and transparency, particularly in data-intensive fields like electrophysiology. Electrophysiology data is typically analyzed using scripts that generate output files, including figures. Handling these results poses several challenges due to the complexity and interactivity of the analysis process. These stem from the difficulty to discern the analysis steps, parameters, and data flow from the results, making knowledge transfer and findability challenging in collaborative settings. Provenance information tracks data lineage and processes applied to it, and provenance capture during the execution of an analysis script can address those challenges. We present Alpaca (Automated Lightweight Provenance Capture), a tool that captures fine-grained provenance information with minimal user intervention when running data analysis pipelines implemented in Python scripts. Alpaca records inputs, outputs, and function parameters and structures information according to the W3C PROV standard. We demonstrate the tool using a realistic use case involving multichannel local field potential recordings of a neurophysiological experiment, highlighting how the tool makes result details known in a standardized manner in order to address the challenges of the analysis process. Ultimately, using Alpaca will help to represent results according to the FAIR principles, which will improve research reproducibility and facilitate sharing the results of data analyses.
\end{abstract}

%
%

\section{Introduction}
\label{sec:intro}

Electrophysiology methods are routinely used to investigate brain function, including the measurement of extracellular potentials using microelectrodes implanted into brain tissue \citep{Huang16_1,Buzsaki12_407}. The first electrophysiology experiments acquired potentials from single or few implanted electrodes, which limited the data throughput of the experiments. However, recent technological advances produced large-density electrode arrays and data acquisition systems able to record hundreds of channels from heterogeneous sources in the experiment sampled at high resolution \citep{Hong19_330}. It is now possible to perform massive and parallel recordings during electrophysiology experiments \citep{Buzsaki04_446} that result in datasets that are both complex in structure and large in volume.

For the analysis of such datasets, this introduces two major consequences. First, the analysis will often be partially conducted in an exploratory style, where the analysis parameters and selection of datasets are probed interactively by the scientists. Keeping track of these choices and approaches is particularly challenging for the scientist in the context of complex data. Second, the analysis of modern datasets often requires advanced methods \citep[e.g.,][]{Brown04_456,Stevenson11_139} that are implemented as workflows composed of several interdependent scripts \citep[see, e.g.,][for a detailed description]{Denker16_58}. The highly diverse and distributed results from the parallel and intertwined processing pipelines operating on complex data must be organized and described in a manner that is comprehensible not only to the original author of the analysis workflow but also in a collaborative context. Taken together, the full workflow including interactive and pipeline approaches, starting from the experimental data acquisition to the presentation of final results, is subject to a hierarchical decision-making process, frequent changes and a large number of processing steps. With growing complexity, these aspects are increasingly difficult to follow, especially in collaborative contexts, where results of analyses executed by different scientists are shared.

The resulting lack of reproducibility undermines the scientific investigations and the public trust in the scientific method and results \citep[cf., e.g.,][]{Baker16_452}. In collaborative environments, the details of an executed analysis workflow should not only be fully documented but also readily understandable by all partners. Thus, work in collaboration could be improved further by directly capturing provenance information on a coarser level of granularity that is informative of the data manipulations throughout the execution of an analysis workflow leading to a certain analysis result \citep{Ragan16_31,Pimentel19_47:1}. By using a provenance tracking system during workflow execution, all operations performed on a given data object can be described and stored in an accessible and structured way that is comprehensible to a human. For the analysis of an electrophysiology dataset, those operations consist of specific analysis methods or processes, such as applying a bandpass filter, downsampling a specific recorded signal, or generating a plot. Ultimately, the details relevant for the final interpretation of the results can be captured and, ideally, stored as metadata with the analysis results. These may then represent summaries of the analysis flow and lead to a description of the results that improve findability, interoperability and reusability of the results \citep[FAIR principles, see][]{Wilkinson16_1}.

Several tools to track and record provenance within (analysis) workflows and single scripts exist, spanning different domains \citep{Ragan16_31,Pimentel19_47:1}. The tools take different approaches depending on which type of information to capture (e.g., tracing code execution, capturing user interactions, or monitoring operating system calls), and the implementation varies according to the intended use of the captured provenance information and its granularity \citep{Bavoil05_135,Murta15_71,Koster12_2520,Pizzi16_218a,MacKenzie-Graham08_208,Davison14_57}. Although some of these solutions might be adapted or even combined to use in the analysis of electrophysiology data, none of these are designed and optimized with the particularities of this type of analysis setting in mind. One of these particularities to consider is the ease of use with custom analysis scripts. A workflow management system (WMS) such as \vistrails \citep{Bavoil05_135}, for instance, requires the construction of workflows from analysis modules implemented as part of the WMS framework or by writing plugins, when the user might need the flexibility and interactivity of custom scripts. Likewise, a tool such as \aiida \citep{Pizzi16_218a} provides a full workflow ecosystem that requires the development of plugins and wrappers to interface and enforces its own data types, hindering the reuse of existing code and libraries without considerable effort. A second aspect to consider is the level of detail and suitability of the provenance information. For individual scripts, tools like \noworkflow \citep{Murta15_71} produce a provenance trail that is highly detailed and without semantic meaning, making it difficult for the scientist to extract information. In contrast, a tool like \sumatra \citep{Davison14_57} will record a more global context in which a script is run in the command line (script parameters, execution environment, version history and links to the output files), while specific operations inside the script will not be detailed. Solutions to orchestrate a series of scripts, like \snakemake \citep{Koster12_2520}, produce flow graphs that show the flow of execution for the scripts composing a workflow but lack the actions performed within each script such that more detailed provenance metadata must be manually recorded by the user without any standardization. Finally, a last aspect is the specificity of the tools for a certain scientific domain. For example, a tool like \loni, that supports full workflows with provenance tracking for the analysis of neuroimaging data \citep{MacKenzie-Graham08_208}, could be readily used in some analysis scenarios. However, the specificity of the available workflow components is a disadvantage for the user that wants to implement pipelines that fall outside of the scope of the intended use.

This work sets out to address the challenges associated with the analysis of an electrophysiology dataset and sharing the results. To accomplish this, a novel tool was implemented to capture the suitable scope of provenance information and store it as metadata together with results generated by analysis scripts implemented in the \python programming language. A typical analysis scenario is presented as a use case and then the tool is analyzed with respect to the challenges it aims to address.

\section{Materials and Methods}

\subsection{Challenges for provenance capture during the analysis of electrophysiology data}

We argue that a tool to capture provenance information during the analysis of electrophysiology data has to deal with four principle scenarios: (i) the analyses often require several preprocessing steps before any analytical method is applied; (ii) the data analysis process is often not linear but intertwined and therefore exhibits a certain level of intricacy; (iii) parameters of the analysis are frequently and often interactively probed; and (iv) the final results are likely to be published or used in shared environments. In the following, we describe these scenarios in detail and derive four associated challenges for capturing provenance.

\begin{figure}
    \begin{fullwidth}
    \centering
    \includegraphics[width=0.95\linewidth]{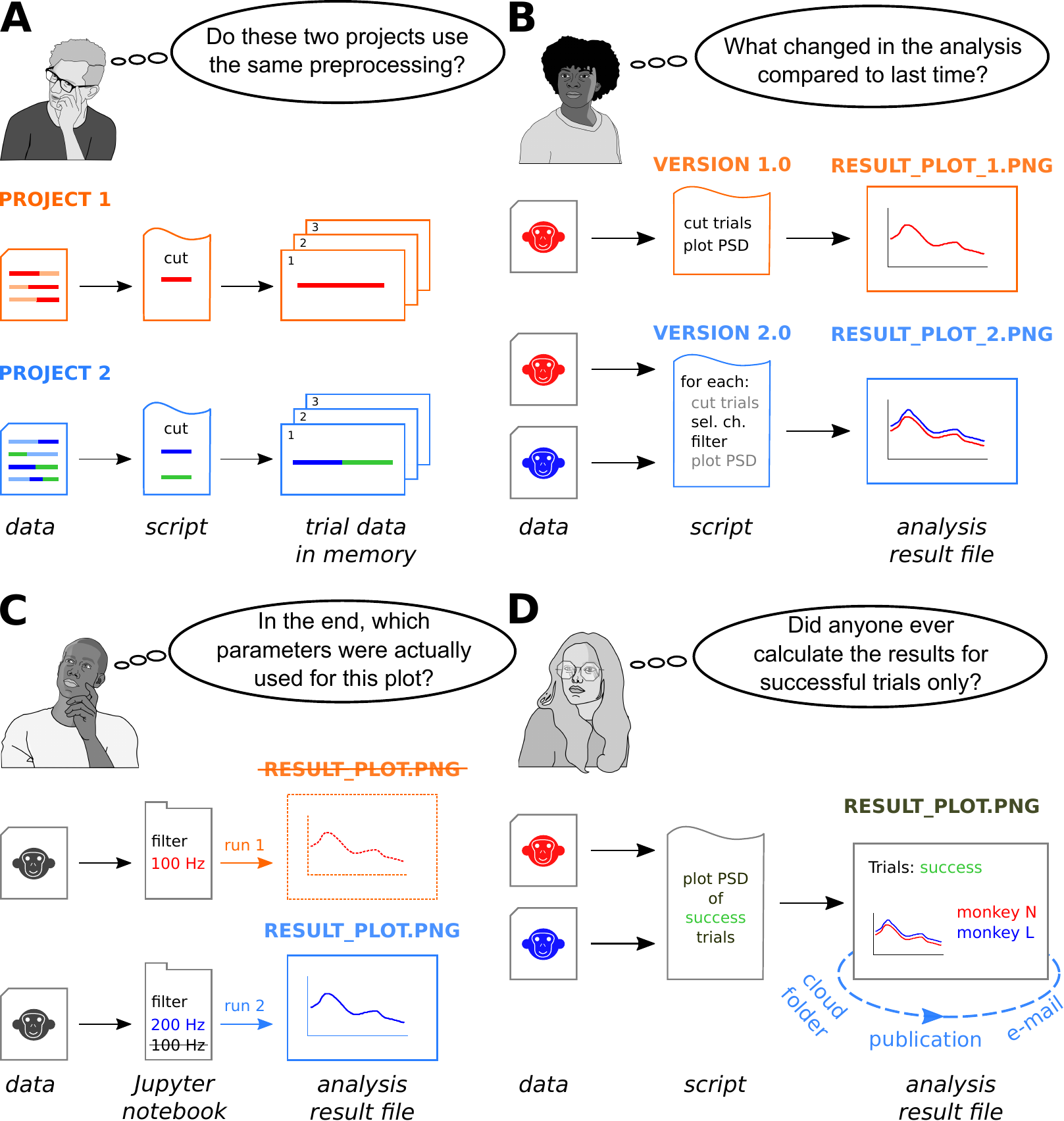}
    \caption{\textbf{Overview of the four representative scenarios leading to challenges associated with the analysis of an electrophysiology dataset. A.} Data frequently requires preprocessing before the analytic methods are applied. This step is often customized for each project, resulting in distinct pipelines that are typically implemented on the level of a script for reasons of efficiency. \textbf{B.} The structure of an analysis script is not static throughout the analysis process. It can be updated to accommodate new data or hypotheses. This results in multiple versions of scripts with increasingly complex pipelines, which are associated with distinct versions of the results. \textbf{C.} Relevant parameters for the analyses are often probed interactively, with the subsequent results changing in time. Keeping track of the exact parameters used for specific results becomes difficult. \textbf{D.} Use of results in shared environments and publishing results requires making available the details of the analysis process to all collaborators. Finding specific results in a shared repository of results from different collaboration partners is difficult, as relevant parameters may be stored in a non-machine-readable format inside the result file, or cryptically in file names.}
    \label{Figure 1}
    \end{fullwidth}
\end{figure}

Preprocessing is a typical step in the analysis, and is usually custom-tailored to a particular project (\cref{Figure 1}A). For instance, data from a recording session of multiple trials (e.g., repeated stimulus presentations or behavioral responses) is usually recorded as a single data stream and only during the analysis cut into the individual trial epochs relevant to the analysis goal. Due to the high level of heterogeneity in the data, this is frequently achieved using custom scripts, with parameters that are specific to the trial structure and design of the experiment (e.g., selecting only particular trials according to behavioral responses such as reaction time). The scientist's written documentation, source code, and, in many cases, the data itself, would need to be inspected to understand all these steps, e.g., the chunking of the data that was performed before the core analysis. Therefore, a first challenge is to clearly document the processing in an accessible and automated manner and to provide this information as supplement to the analysis output.

The full analysis pipeline from the dataset to a final result artefact is likely not built in one attempt, but instead involves a continuous development (\cref{Figure 1}B). For instance, as new data is obtained, time series may need to be excluded from analysis and new hypotheses are generated. Therefore, the analysis scripts may be updated to include additional analysis steps, and the resulting code will have increasing complexity. One solution to organize this agile process is to use a WMS (e.g., \snakemake; \cite{Koster12_2520}) coupled with a code versioning system such as \git. For each run, the WMS will provide coarse provenance information, such as the name of the script, environment information, script parameters, and files that were used or generated. The scripts can then be tracked to specific versions knowing the \git commit history. However, if multiple operations (e.g., cutting data, downsampling and filtering) are performed inside one script, the actual parameters in each step are possibly not captured as part of the provenance. This is the case where provenance information shows only script parameters passed by command line. The mapping of command-line to the actual parameters used by the functions in the script relies on the correct implementation of the code, and any default parameters for the function that are not passed by command-line will not be known. Furthermore, it is not possible to inspect each intermediate (in-memory) data object during the execution of the script. Yet, without knowledge of these data operations and the data flow it becomes challenging to compare results generated by multiple versions of the evolving analysis script, in particular if the code structure of the script changes over time. A solution to this challenge could be to break such complex scripts in several smaller scripts, such that the coarse provenance information of the WMS could be more descriptive of each individual process and intermediate results would be saved to disk (i.e., in our example, separate scripts for cutting, downsampling and filtering). However, this may be inconvenient and inefficient: resource-intensive operations (e.g., file loading and writing) might be repeated across different scripts, and temporary files would have to be used between the steps, instead of efficiently manipulating data in memory. Moreover, this approach limits the expressiveness and creativity of defining data operations as opposed to the full set of operations offered by the programming language in a single script. Therefore, a second challenge is to efficiently capture  the parameters and the data flow associated with the analysis steps of the script. 

The parameters that control the final analysis output are frequently probed interactively (\cref{Figure 1}C). For example, the scientist performing the analysis could write a \jupyter notebook \citep{Kluyver16_87} to find specific frequency cutoffs for a filtering step. In one scenario, code cells of the notebook can be run in arbitrary sequences, with some parameters being changed in the process until a result artifact (e.g., a plot) is saved in a file. In a different scenario, it is possible to generate several versions of a given file by the same notebook, each of which overwrites the previous version. At this point, the scientist performing the analysis might rely in the associated \jupyter history or versioning of the notebook/files using \git. However, the relevant parameters that were used to generate results saved in last version of the file would be difficult to recall. Ultimately, a detailed documentation by the user or retracing the source code according to an execution history is still required. Therefore, a third challenge is to retain a documentation of the interactive generation of the analysis result that is explicitly and unambiguously linked to the generated result file.

The fourth challenge stems from the situation where results (e.g., plots) are likely to be published or used in collaborative environments (\cref{Figure 1}D). This includes files uploaded in a manuscript submission, or files deposited in a shared folder or sent via e-mail between collaborators. The interpretation of the stored results depends on the understanding of the analysis details and its relevant parameters by the collaboration partner. Moreover, searching for specific results in a large collection of shared files can be difficult: not all the relevant parameters are recorded in the file name, and are likely stored as non machine-readable information within the file (e.g., an axis label in a figure). In these situations, analysis provenance stored together with the shared result files as structured and comprehensible metadata should improve information transfer in the collaboration and findability of the results.

\subsection{Use case scenario}
\label{sec:use-case}

\begin{figure}
    \begin{fullwidth}
    \centering
    \includegraphics[width=0.90\linewidth]{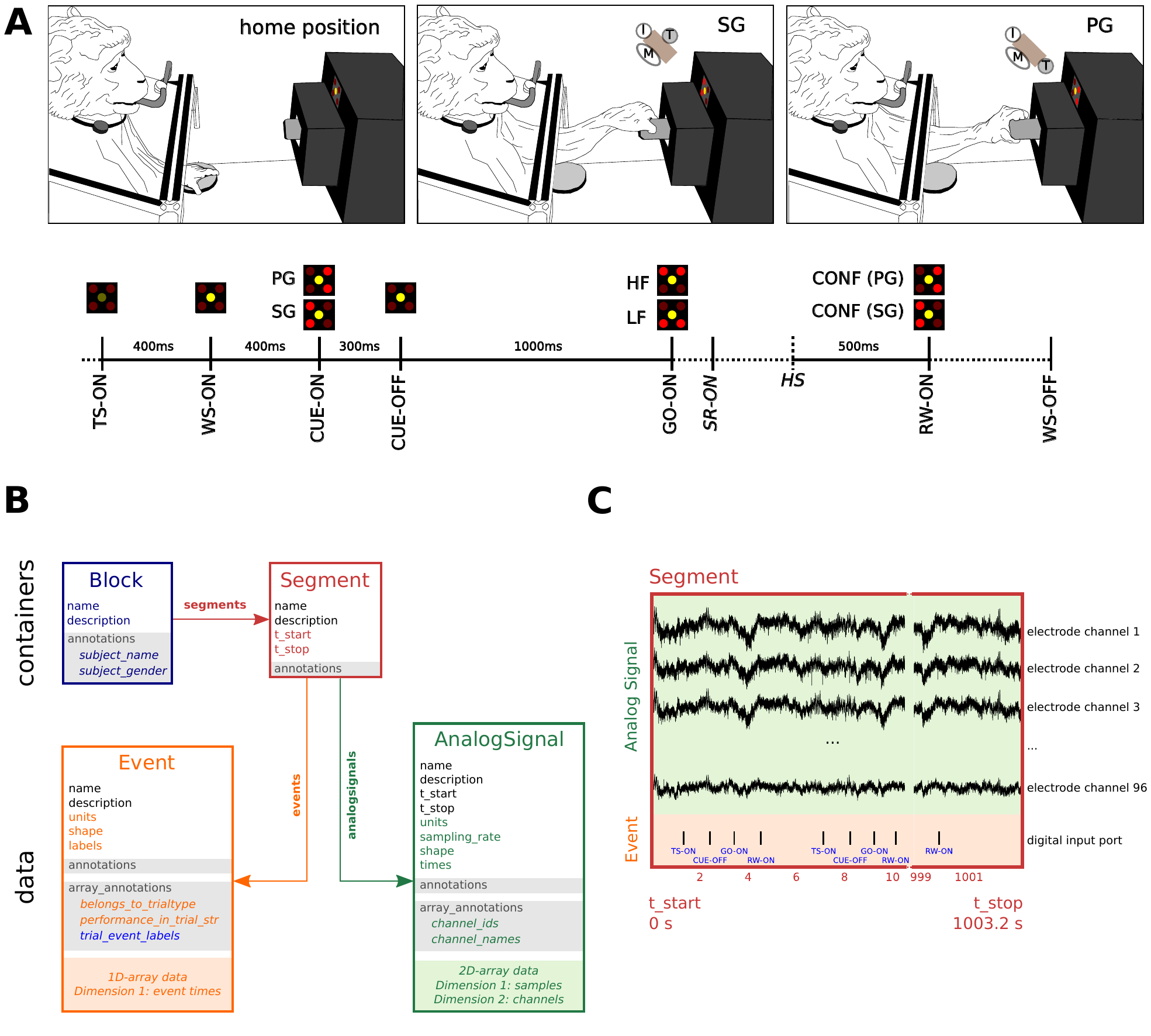}
    \caption{\textbf{Overview of the delayed reach-to-grasp task and \neo data model.} The datasets from the experiment were used for use case scenario for capturing provenance during the analysis of electrophysiology data, and \neo was used to load and manipulate the data in the analysis script. \textbf{A.} Description of the experimental protocol (cf.~ \cite{Brochier18_180055} for details). The monkey is instructed to grasp an object using either a side grip (SG) or precision grip (PG), and this is identified as the CUE-ON event. After a 1~s delay, a GO signal (GO-ON event) marks the start of the movement and indicates whether to pull the object with either low (LF) or high (HF) force. Several behavioral events occur during a trial (marked on the time line). \textbf{B.} Overview of the \neo data model defining container objects (shown here: \neoBlock and \neoSegment) and data objects (shown here: \neoAnalogSignal and \neoEvent). Supporting metadata are stored as \textit{annotations} and \textit{array\_annotations} dictionaries (gray shading). Annotations are single values associated with a key (e.g., \textit{subject\_name}). Array annotations are stored in arrays with the length of the data (e.g., number of events in \neoEvent or number of channels in \neoAnalogSignal). \neoSegment objects group data objects in specific windows of time (given by attributes \textit{t\_start} and \textit{t\_stop}). \neoEvent objects are arrays with timestamps of labeled events. \neoAnalogSignal objects are two-dimensional arrays (first dimension: time, second dimension: channels) where the time axis is determined by the \textit{t\_start} and \textit{sampling\_rate} attributes. The \textit{units} attribute identifies the physical unit associated with the data array (e.g., microvolts for \neoAnalogSignal and seconds for \neoEvent). \textbf{C.} Time-homogeneous representation of two data elements of the dataset (\neoAnalogSignal and \neoEvent) stored inside a \neoSegment container. Selected timestamps and corresponding values of the array annotation \textit{trial\_event\_labels} of the \neoEvent object are presented in blue. Image in panel A is adapted from \cite{Brochier18_180055}, licensed under the Creative Commons Attribution 4.0 International License.}
    \label{Figure 2}
    \end{fullwidth}
\end{figure}

As a use case scenario, we consider an analysis that computed the mean power spectral densities (PSDs) from a publicly available dataset containing massively parallel electrophysiological recordings (raw electrode signals, local field potentials, and spiking activity) in the motor cortex of monkeys in a behavioural task involving movement planning and execution. The experiment details, data acquisition setup, and resulting datasets were previously described \citep{Brochier18_180055}. Briefly, two subjects (monkey N and monkey L) were implanted with one Utah electrode array (96 active electrodes) in the primary motor/premotor cortices. Subjects were trained in an instructed delayed reach-to-grasp task. In a trial, the monkey had to grasp a cubic object using either a side grip (SG) or a precision grip (PG). The SG consists of the subject grasping the object with the tip of the thumb and the lateral surface of the other fingers, on the lateral sides of the object. The PG consists of the subject placing the tips of the thumb and index finger on a groove on the upper and lower sides of the object. The monkey had to pull the object against a load that required either a low (LF) of high pulling force (HF). The grip and force instructions were presented through an LED panel using two different visual cue signals (CUE and GO), respectively, which were separated by a 1000~ms delay (\cref{Figure 2}A). As a result of the combination of the grip and force conditions, four trial types were possible: SGLF, SGHF, PGLF, and PGHF. A recording session consisted of several repetitions of each trial type that were acquired continuously in a single recording file. Neural activity was recorded during the session using a Blackrock Microsystems Cerebus data acquisition system, with the raw electrode signals bandpass-filtered between 0.3 and 7500~Hz at the headstage level and digitized at 30~KHz with 16-bit resolution (0.25~V/bit, raw signal). The behavioral events were simultaneously acquired through the digital input port that stored 8-bit binary codes as received from the behavioral apparatus controller. 

The experimental datasets are provided in the \textit{Neuroscience Information Exchange} (\nix\footnote{RRID:SCR\_016196; \url{https://nixio.readthedocs.io}}) format, developed with the aim to provide standardized methods and models for storing neuroscience data together with their metadata \citep{Stoewer14}. Inside the \nix file, data are represented according to the data model provided by the \neo\footnote{RRID:SCR\_000634; \url{https://neuralensemble.org/neo}} \python library \citep{Garcia14_10}. \neo provides several features to work with electrophysiology data. First, it allows loading data files written using open standards such as \nix as well as proprietary formats produced by specific recording systems (e.g., Blackrock Microsystems, Plexon, Neuralynx, among others). Second, it implements a data model to load and structure information generated by the electrophysiology experiment in a standardized representation. This includes time series of data acquired continuously in samples (such as the signals from electrodes or analog outputs of a behavioral apparatus) or timestamps (such as spikes in an electrode or digital events produced by a behavioral apparatus). Third, \neo provides typical manipulations and transformations of the data, such as downsampling the signal from electrodes or extracting parts of the data at specific recording intervals. The objects may store relevant metadata, such as names of signal sources, channel labels, or details on the experimental protocol. In this use case scenario, \neo was used to load the datasets and manipulate the data during the analysis. 

The relevant parts of the structure and relationships between objects of the \neo data model are briefly represented in \cref{Figure 2}B. The \neo library is based on two types of objects: data and containers. Different classes of data objects exist, depending on the specific information to be stored. Data objects are derived from \quantity arrays that are provided by the \python \quantities package\footnote{\url{https://github.com/python-quantities/python-quantities}} and provide \numpy arrays with attached physical units. The \neoAnalogSignal is used to store one or more continuous signals (i.e., time series) sampled at a fixed rate, such as the 30~kHz raw signal captured from each of the 96 electrodes in the Utah array. The \neoEvent object is used to store one or multiple labeled timestamps, such as the behavioral events throughout the trials acquired from the digital port of the recording system. The container objects are used to group data objects together, and these are accessed through specific collections (lists) present in the container. The top-level container is the \neoBlock object that stores general descriptions of the data and has one or more \neoSegment objects accessible by the \textit{segments} attribute. The \neoSegment object groups data objects that share a common time axis (i.e., they start and end within the same recording time, defined by the \textit{t\_start} and \textit{t\_stop} attributes; \cref{Figure 2}C). The \neoSegment object also has collections to store specific data objects: \textit{analogsignals} is a list of the \neoAnalogSignal data objects, and \textit{events} is a list of the \neoEvent data objects. 

The \neo data model also defines a framework for metadata description as key-value pairs for its data and container objects through annotations and array annotations. Annotations may be added to any \neo object. They contain information that are applicable to the complete object, such as the hardware filter settings that apply to all channels contained in an \neoAnalogSignal object. Array annotations may be added to \neo data objects only. They contain information stored in arrays, whose length corresponds to the number of elements in the data. They are used to provide metadata for a particular element in the data stored in the object. For instance, in the \neoEvent object representing the behavioral events in the reach-to-grasp task, the \textit{trial\_event\_labels} array annotation stores the decoded event string associated with each event timestamp stored in the object (\cref{Figure 2}C). In the end, all the data in the \nix dataset is loaded into \neo data objects that encapsulate all the relevant metadata.

\begin{figure}
    \begin{fullwidth}
	\centering{}
	\includegraphics[width=\linewidth]{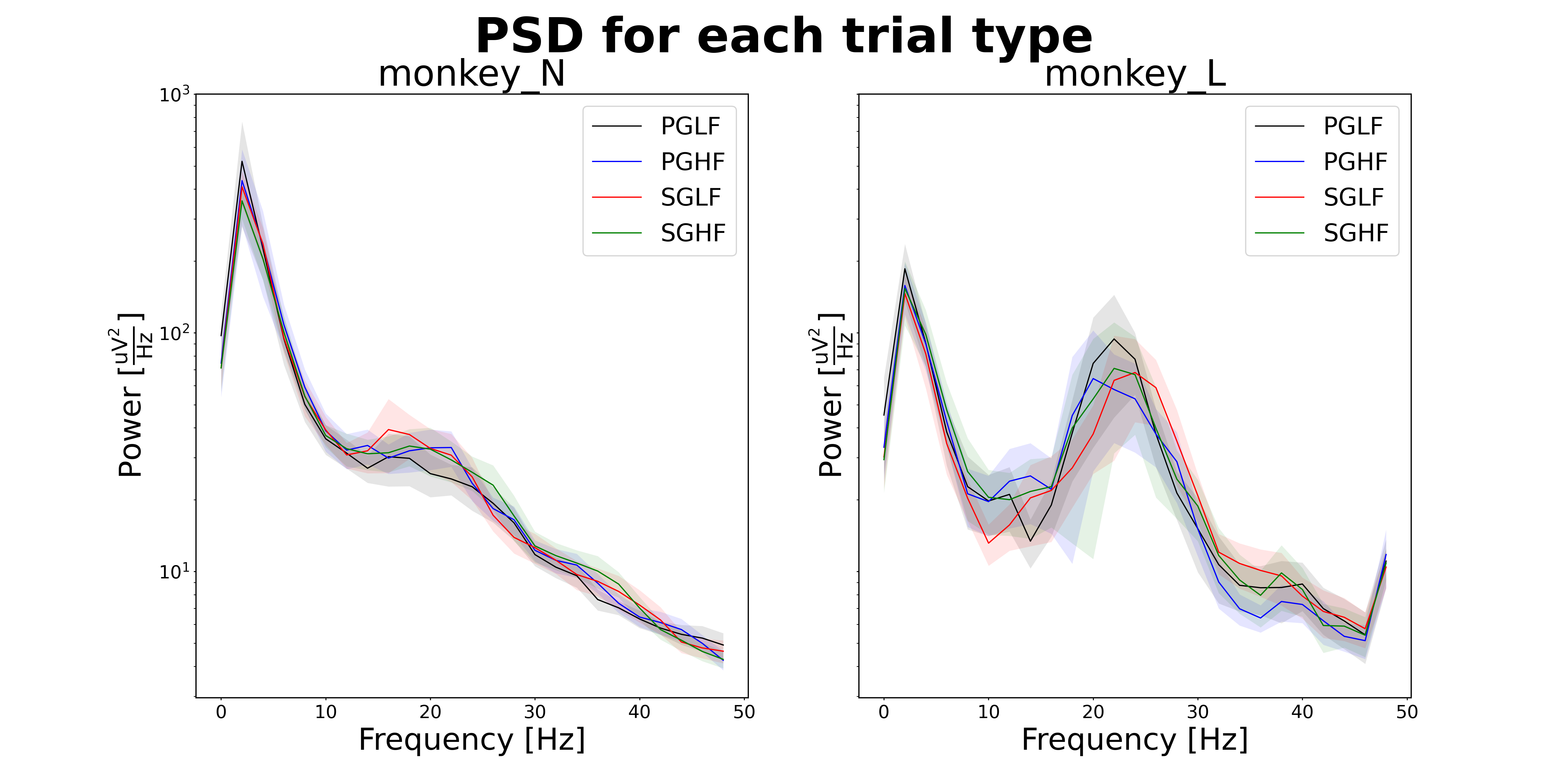}
	\caption{\textbf{Output of the analysis workflow implemented in the script \scriptfile in the context of the use case scenario}. This plot was stored as a PNG file named \plotfile.}
    \label{Figure 3}
    \end{fullwidth}
\end{figure}

\begin{figure}
    \begin{fullwidth}
    \centering
    \includegraphics{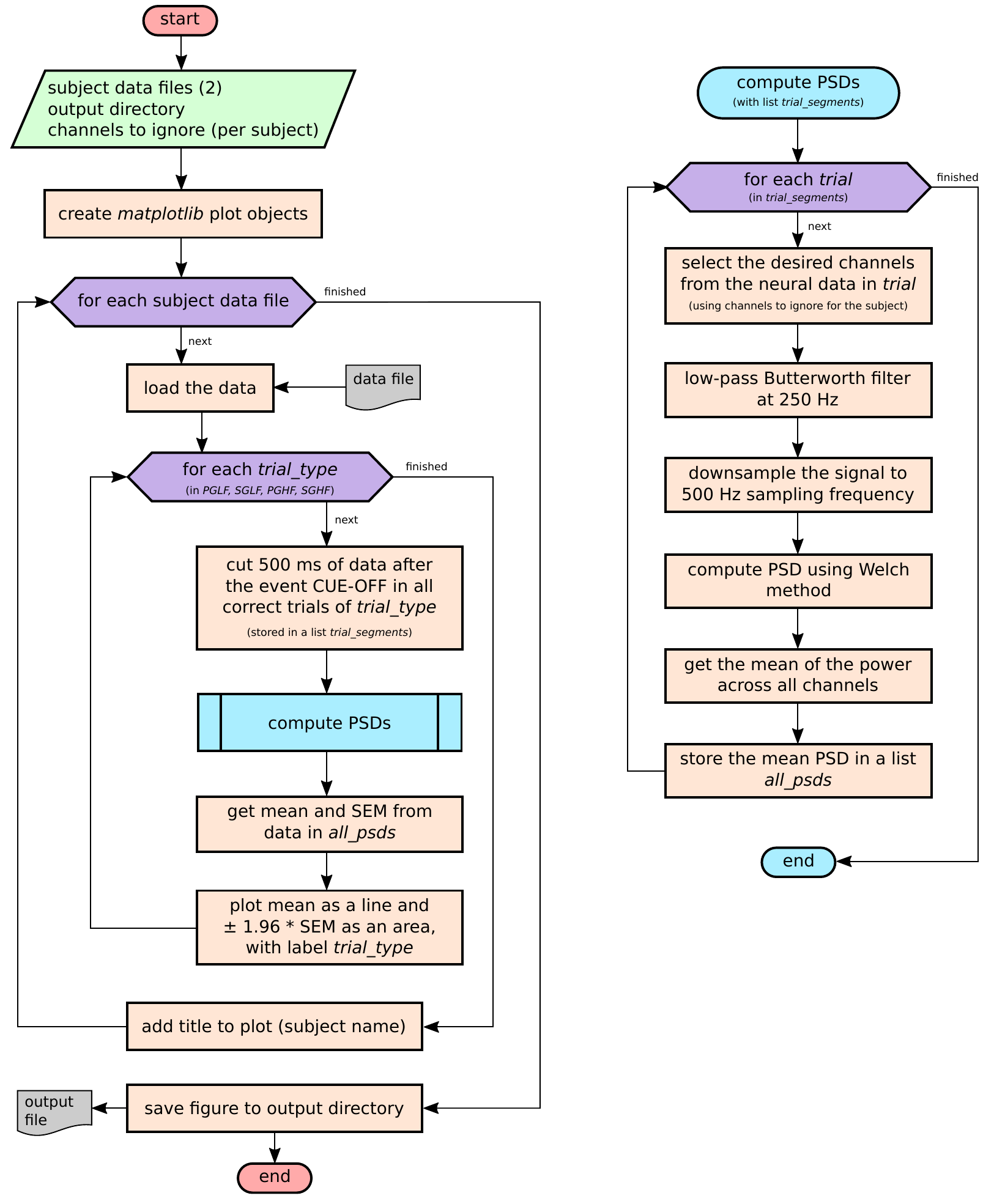}
    \caption{\textbf{Flowchart of the analysis implemented as use case scenario.} The code for this algorithm is implemented in \scriptfile. The main steps are composed by three nested loops (purple hexagons): for each input file, a second loop runs over the 4 possible trial types, extracting the data of the individual trial epochs. After that, for each trial, the channel-wise PSDs are computed and the power density estimates from all channels are subsequently aggregated. At the end of each trial type loop, the single-trial PSDs are aggregated and plotted. After the last input file is processed, the plot is saved to \plotfile.}
    \label{Figure 4}
    \end{fullwidth}
\end{figure}

In the use case scenario, the PSDs were analyzed for each subject (monkey N and monkey L), and the mean PSD was computed for each of the four trial types present in the experiment (\cref{Figure 3}). Although a single \python script (named \scriptfile) was used to produce the plot (stored as \plotfile), the actual analysis algorithm is complex (shown in a schematic form in \cref{Figure 4}). In a typical scenario, a file such as \plotfile could be stored in a shared folder or even sent to collaborators by e-mail. At this point, several key information cannot be obtained from the plot alone: (i) How were the trials defined, i.e., which time points or behavioral events were used as start and end points to cut the data in the data preprocessing? (ii) Was any filtering applied to the raw signal, before the computation of the PSD? (iii) Several methods are available to obtain the PSD estimate, each with particular features that may affect the estimation of the spectrum (see \cite{Welch67_70} and \cite{Percival93}). Which method was used in this analysis, and what were the relevant parameters (e.g., for frequency resolution)? (iv) How was the aggregation performed (i.e., method and number of trials). What do the shaded area intervals around the plot lines represent? In addition to these questions, the contents of a plot such \plotfile may be the result of several iterations of exploratory analyses and development of \scriptfile. In our scenario, parameters that could have been iteratively probed or improved could be the identification of failed electrodes, definition of a suitable time window for cutting the data from a full trial, or to select specific filter cutoffs. Therefore, \plotfile could be overwritten after \scriptfile was run with different parameters or different versions of the code. Altogether, the exhaustive set of steps and definitions used for the generation of the analysis result is not apparent from \plotfile. Even with a good description such as the flowchart in \cref{Figure 4}, that could be added as accompanying documentation, the exact parameters used for function calls are still missing, especially if these were determined during run-time (such as the number of trials in the dataset).

\begin{figure}
    \begin{fullwidth}
    \centering
    \includegraphics[scale=0.8]{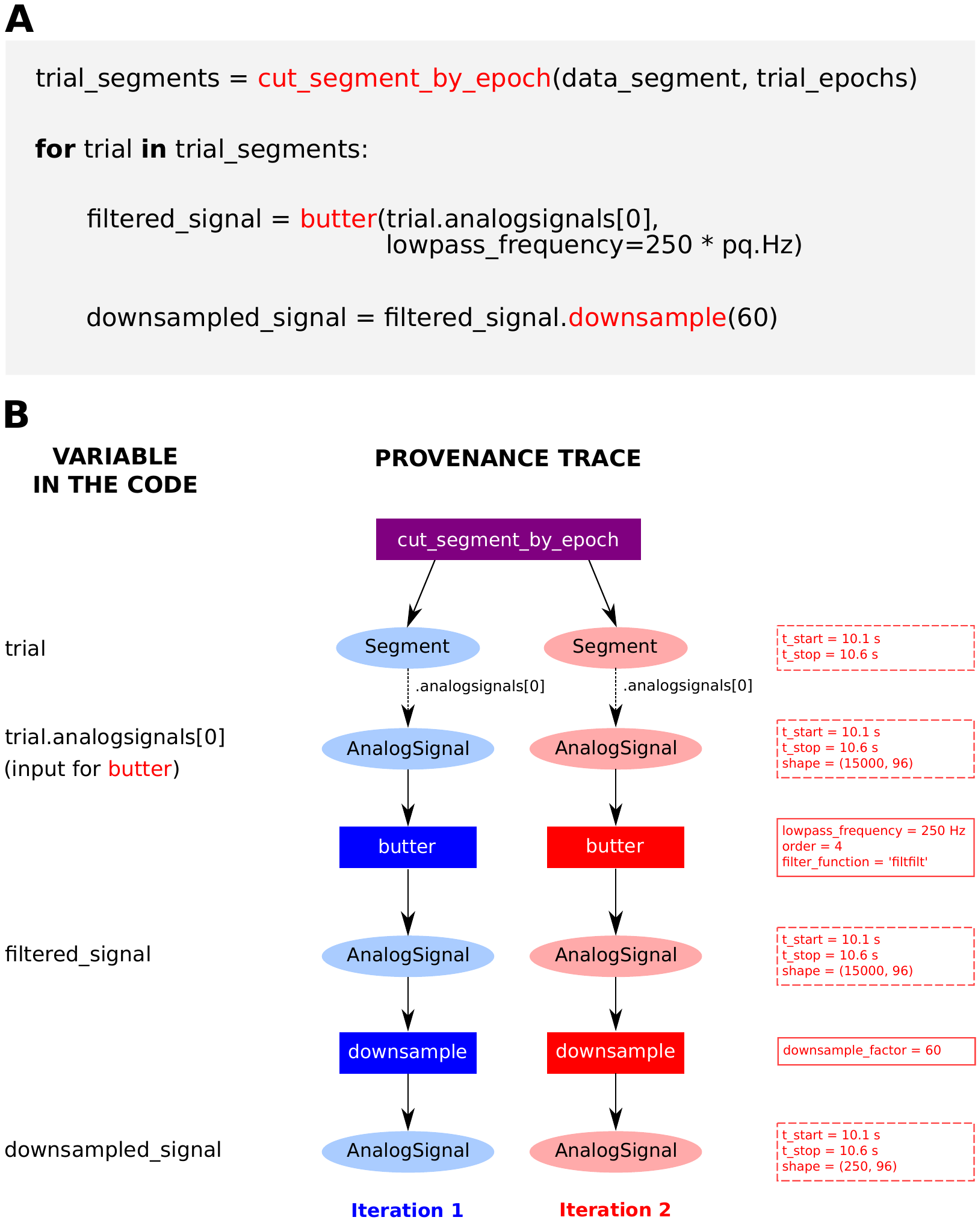}
    \caption{\textbf{Provenance helps to understand exact data processing during code execution. A.} Snippet of code in \python that iterates over a list with individual trial data to perform a low-pass filter operation using a Butterworth filter followed by downsampling the signal by a factor of 60. In this example, the code uses data objects defined by the \neo framework \citep{Garcia14_10}. The list \textit{trial\_segments} stores several \neoSegment objects. The raw neural signal from the electrode array is stored in \neoAnalogSignal objects. Note that although knowing the hierarchical structure of \neo helps to understand the implemented code, the actual data objects and any associated information cannot be accessed, as these exist only during run time. \textbf{B.} Example of a provenance trace to represent the execution of the code in A. Data objects are represented as ellipses, and functions as rectangles. Two exemplary iterations of the loop are shown as two separate paths in a graph (highlighted by blue or red color shades, respectively). Dashed lines represent accessing a data object contained into another data object using a specific \python operation (e.g., subscript or attribute). Note that the \neoSegment object stored in the variable \textit{trial} is known for every single iteration, and its transformations are followed individually. The information associated with each data object (e.g., start and end time of \neoSegment, or the shapes of the \neoAnalogSignal objects) can be inspected at every stage of the iteration (dashed red rectangles on the right show information for the second loop iteration). Parameters of the functions called, even if they were not explicitly written in the code, but determined during run-time, can also be inspected (solid line red rectangles on the right show parameters for each function called in the second iteration).}
    \label{Figure 5}
    \end{fullwidth}
\end{figure}

In the end, the only way of getting those relevant details of the analysis is by directly inspecting \scriptfile. The difficulties associated with this approach are illustrated in \cref{Figure 5}. For a simple code snippet (\cref{Figure 5}A), which iterates over a list of trial data to apply a Butterworth filter and then downsample the signal, it is not possible to visualize the state of the data for each iteration (e.g., the array shape). In addition, the actual contents of the variables are unknown. A robust data model like \neo helps to understand which objects were accessed during each iteration. However, even when using that framework, the exact data objects and their transformations in each iteration of the for-loop are not apparent from the code given that the object instances (including attributes, such as the shape of an array) are only available during run time. One example of such information that exists only at run time is the number of trials (i.e., the number of \neoSegment objects returned by \textit{cut\_segment\_by\_epoch}) and the number of channels (i.e., the shape of the \neoAnalogSignal object in each loop iteration). Unless running the script again with the same dataset and explicitly outputting this information, it is not possible to know. In contrast, by capturing and structuring the relevant provenance during the execution, a representation could be obtained in a way that all relevant information is accessible after the run (\cref{Figure 5}B). The detailed trace ultimately shows which part of the data and the resulting intermediate objects were used during each iteration.

\subsection{Alpaca: a tool for automatic and lightweight provenance capture in \python scripts}

As the analysis of electrophysiology datasets is usually based on scripts such as \scriptfile, we set to implement Alpaca (Automated Lightweight ProvenAnce CApture) as a tool to capture the provenance information that describes the main steps implemented in scripts that process data. The captured information can be stored as a metadata file that is associated with the result file(s) generated by the script (e.g., the plot in \cref{Figure 3} stored in \plotfile). Alpaca can be used for scripts written in the \python programming language as \python is free and open source, and has been gaining popularity among the neuroscience community \citep{Muller15_11}. \python is also frequently used in the analysis of electrophysiology data, and several dedicated open source packages are available, such as the \neo and \textit{NWB}\footnote{Neurodata Without Borders; RRID:SCR\_015242; \url{https://www.nwb.org}} \citep{Ruebel22_e78362} frameworks for electrophysiology data representation, the unified spike sorting pipeline \textit{SpikeInterface}\footnote{RRID:SCR\_021150; \url{https://spikeinterface.readthedocs.io}} \citep{Buccino20_e61834}, and \elephant\footnote{Electrophysiology Analysis Toolkit; RRID:RRID:SCR\_003833;  \url{https://python-elephant.org}} \citep{Denker18_P19} for data analysis. Therefore, a tool implemented in \python will have greater impact in the neuroscience community, as no licenses or fees are required and it builds on already established state-of-the-art processing and analysis tools.

\begin{figure}
    \begin{fullwidth}
    \centering
    \includegraphics[width=0.8\linewidth]{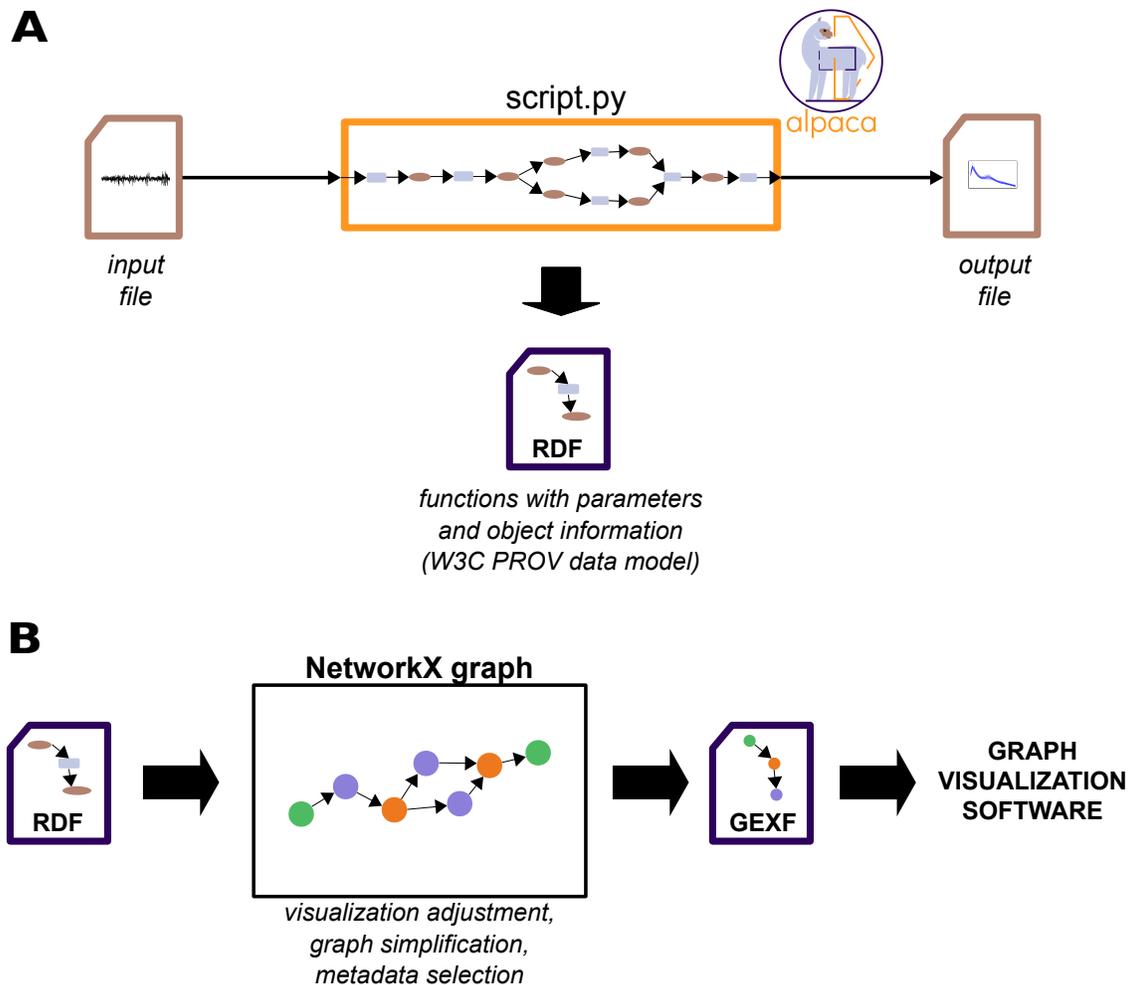}
    \caption{\textbf{Schematic overview of the functionality of Alpaca. A.} The decorator and functions provided by Alpaca are incorporated into a \python script that processes data (orange rectangle). The script reads an input file and generates another file as result output. Alpaca tracks the functions called during the execution of the script (represented by the light blue rectangles inside the orange rectangle) together with the input and output data objects (represented by the brown ellipses within the orange rectangle). Function parameters and metadata of data objects are also captured. The aggregated information is structured according to the W3C PROV standard and serialized to a file (dark blue) using RDF acting as a sidecar file to the output file of the script. \textbf{B.} To visualize the captured provenance, the serialized RDF files can be converted into \networkx graph objects using Alpaca. The graph can be adjusted to concentrate on specific information of the captures provenance, or simplified. Finally, the graph can be saved using graph serialization formats (e.g., GEXF) supported by third-party graph visualization software (e.g., \gephi).}
    \label{Figure 6}
    \end{fullwidth}
\end{figure}

The functionality of Alpaca is illustrated in \cref{Figure 6}. Alpaca is based on a \python function decorator\footnote{A \python decorator allows adding new functionality to existing functions without changing their behavior.} that supports tracking the individual steps of the analysis and constructing a provenance trace. In addition, Alpaca serializes the captured provenance information (\cref{Figure 6}A) as a metadata file encoded in the RDF\footnote{Resource Description Framework, a general model for description and exchange of graph data.} format \citep{RDF_CORE} according to the data model defined in the W3C\footnote{World Wide Web Consortium, \url{https://www.w3.org}} PROV standard (PROV-DM; \cite{W3C_PROV_DM}). PROV is an open standard that was developed to allow the interoperability of provenance information in heterogeneous environments \citep{W3C_PROV_OVERVIEW}. Finally, visualization of the provenance trace is supported by converting the PROV metadata into graphs that show the data flow within the script and allow the visual inspection of the captured provenance (\cref{Figure 6}B). Alpaca is provided as a standalone \python package that can be installed from the \textit{Python Package Index} or directly from the code repository\footnote{\url{https://github.com/INM-6/alpaca}}. The documentation with usage examples is available online\footnote{\url{https://alpaca-prov.readthedocs.io}}. 

Several design decisions were adopted in Alpaca. First, the tool captures provenance during the execution without the need for users to enhance this information with additional metadata or documentation. Second, code instrumentation is reduced to a minimum level, and users are asked to make only minor changes in the existing code to enable tracking (see Suppl.~Text 1 for the changes required to track provenance within \scriptfile). Third, it is flexible enough to accommodate different coding styles, and it was designed to be the most compatible with existing code bases. Therefore, provenance is captured in an automatized and lightweight fashion.

Alpaca assumes that an analysis script such as \scriptfile is composed of several functions that are called sequentially (potentially in the context of control flow statements such as loops), each performing a step in the analysis. The functions in the script may take data as input and produce outputs based on a transformation of that data, or generate new data. Moreover, a function may have one or more parameters, that are not data inputs but modify the behavior on how the function is generating the output. For example, in reshaping an array using the \numpy function \textit{reshape}, the new shape would represent a parameter that defines how to reshape the original array (i.e., input data) into a new array (i.e., the output data). In \python, information to a function is passed through function arguments, that are accessed by the local code in the function body that performs the computation. Those are specified in the function declaration using the \textit{def} keyword. Therefore, Alpaca utilizes the following definitions to analyze a function call in the script:

\begin{itemize}
	\item \textbf{input:} a file or \python object that provides data for the function. It is one of the function arguments;
	\item \textbf{output:} a file or \python object generated by a function. Can be a return value of the function or one of the function arguments;
	\item \textbf{parameter:} any other function argument that is neither an input nor an output;
	\item \textbf{metadata:} additional information contained in the input/output. For \python objects, these can be accessible by attributes (i.e., accessed by the dot . after the object name, such as \textit{signal.shape}) or annotations stored in dictionaries accessed by special attributes, such the ones defined in the \neo data model. For files, this is the file path.
\end{itemize}

\subsubsection{Initializing Alpaca}

The calls to the functions tracked by Alpaca are expected to be present in a single scope (i.e., the main script body or a single function such as \textit{main}). To identify the code to be tracked and start the capture, the user must insert a call to the \alpacaActivate function at a point in the script before the corresponding block of code. When calling \alpacaActivate, Alpaca identifies the current script in execution, obtains the SHA256 hash\footnote{A hash is a function that maps data with variable size to fixed-size values. SHA256 is a Secure Hash Algorithm (SHA) that can be used to verify the identity of files.} of the source file storing the code, and generates an UUID (Universally Unique Identifier) to uniquely identify the script execution (\alpacaSessionID). The source code to be tracked will be analyzed to allow the extraction of each individual code statement later, during the analysis of each function execution.

Before activating the tracking, the user can set options using the \alpacaSettings function. These settings operate globally within the toolbox and control how Alpaca captures and describes provenance.

\subsubsection{Tracking the steps of the analysis}

\begin{figure}
    \begin{fullwidth}
    \centering
    \includegraphics[width=\linewidth]{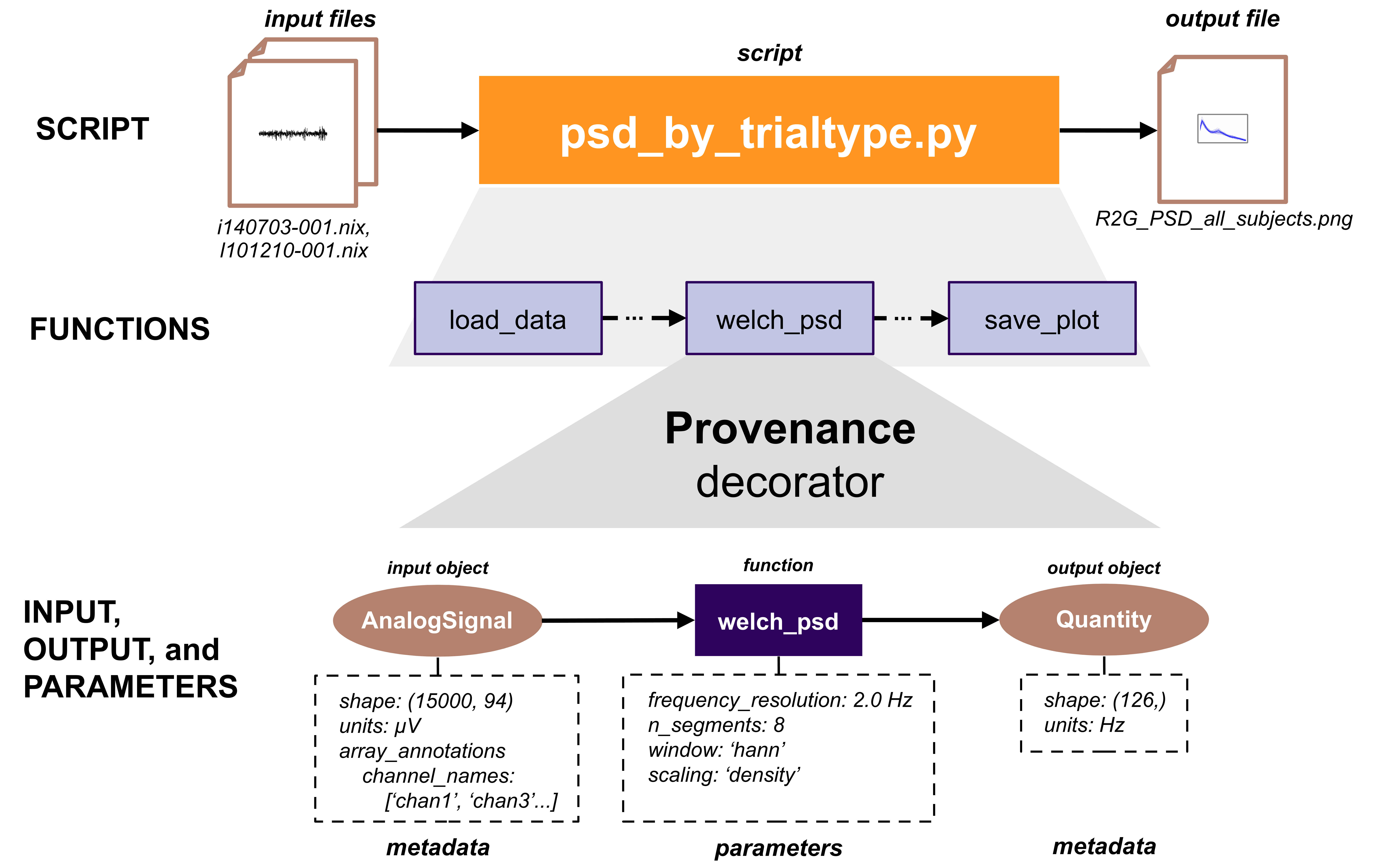}
    \caption{\textbf{Alpaca captures fine-grained provenance information at each step during the execution of \python scripts that process data.} The \alpacaProvenance decorator is used to wrap each function called in the script. The input and output data objects for each function are identified, and any embedded object metadata is captured (bottom right and left). Object metadata are attributes of the objects or special values stored in annotation dictionaries. This takes advantage of available data models for electrophysiology where experimental details can be stored together with the data as annotations (e.g., \neo). In the example provided in this paper, the object metadata are attributes such as shape and units, together with annotations such as channel names in an \neoAnalogSignal with the data recorded from the electrodes. Finally, the parameters of the functions are also identified (bottom middle). In this example, this is the type of window or the number of segments used in the computation of the PSD using the \elephantWelch function, that implements the Welch algorithm. In the end, a full provenance graph with the lineage from the input files (such as the two \nix files in the example) to the output file with the analysis result (such as \plotfile) is produced.}
    \label{Figure 7}
    \end{fullwidth}
\end{figure}

The \alpacaProvenance function decorator is used to wrap each data processing function executed in the script (\cref{Figure 7}). When applying the decorator, the argument names that are either \python object inputs, file inputs, or file outputs are identified through the decorator constructor parameters \textbf{inputs}, \textbf{file\_input}, or \textbf{file\_output}. When the script is run, for each execution of the function, the decorator: (i) generates a description of the inputs and outputs; (ii) records the parameters used in the call; (iii) generates a unique execution UUID (\alpacaExecutionID); and (iv) captures the start/end timestamps. Finally, this information is used to build a record for the function execution. \alpacaProvenance has an internal global function execution counter, incremented after the execution of any function being tracked. The current value is also added to the function execution record, to obtain the order of that execution. Finally, all the execution records are stored in an internal history, which will be used to serialize the information at the end.

The \alpacaProvenance decorator analyzes the inputs and outputs to extract the information relevant for their description and their metadata:

\begin{itemize}
	\item for \python objects (e.g., an \neoAnalogSignal object), the type information (\python class name and the module where it is implemented), content hash, and current memory address is recorded. The content hash is computed using either the \hash function from the \joblib\footnote{\url{https://joblib.readthedocs.io}} package (using the SHA1 algorithm) or the builtin \python \hash function (that uses the algorithm implemented in the \textit{\_\_hash\_\_} method of the object). By default, every object will be hashed using \joblib. However, it is possible to define specific packages whose objects will be hashed using the builtin \hash function using the \alpacaSettings function. This allows selecting hashing functionality that may already be implemented in the object (which can be faster), or avoid sensitivity to minor changes to the object content that will produce a provenance trace that is too detailed. The values of all object instance attributes (i.e., stored in the \textit{\_\_dict\_\_} dictionary) are recorded, together with the values of the specific attributes when present. This includes, for example, \textit{shape} and \textit{dtype} for \numpy arrays, or extended attributes such as \textit{units}, \textit{t\_start}, \textit{t\_stop}, \textit{nix\_name}, and \textit{dimensionality} for the \neoAnalogSignal object of \neo representing a measurement time series. More generic attributes that could be used by other data models, such as \textit{id}, \textit{pid}, or \textit{create\_time} are also captured if present;
	
	\item for files, the SHA256 file hash is computed using the \hashlib package, and the absolute file path is recorded; 
	
	\item for the \python builtin \textit{None}, the object hash is an UUID, as it is a special case where the actual object is shared throughout the execution environment. This avoids duplication. 
\end{itemize} 

The information on the function is also extracted: name, module, and version of the package where it was implemented (if available through the \textit{metadata} module from the \textit{importlib} package implemented in \python 3.8 or higher). Version information is currently not recorded for user-defined functions (i.e., implemented in the script file being tracked).

Finally, the inputs to a function may be accessed from container objects by subscripts (e.g., an item in a list such as \textit{signals[0]}) or attributes (e.g., \textit{segment.analogsignals}). To capture these static relationships, the abstract syntax tree of the source code statement containing the current function call is analyzed, all container objects are identified, and the operations (subscript or attribute) are added to the execution history. In the end, the container memberships are identified and recorded if used when passing inputs to a function.

\subsubsection{Serialization of the provenance information}

The captured provenance is serialized as RDF graph \citep{RDF_CORE}, using one of the formats supported by \rdflib\footnote{\url{https://github.com/RDFLib/rdflib}}. The \alpacaAlpacaProvDocument class is responsible for managing the serialization, based on the history captured by the \alpacaProvenance decorator. For simplified usage, the serialization can be accomplished in a single step by just calling the \alpacaSaveProvenance function at the end of the script execution, passing a destination file and serialization format. All the information currently stored in the history in \alpacaProvenance will be saved to the disk.

\begin{figure}
    \begin{fullwidth}
    \centering
    \includegraphics[width=0.85\linewidth]{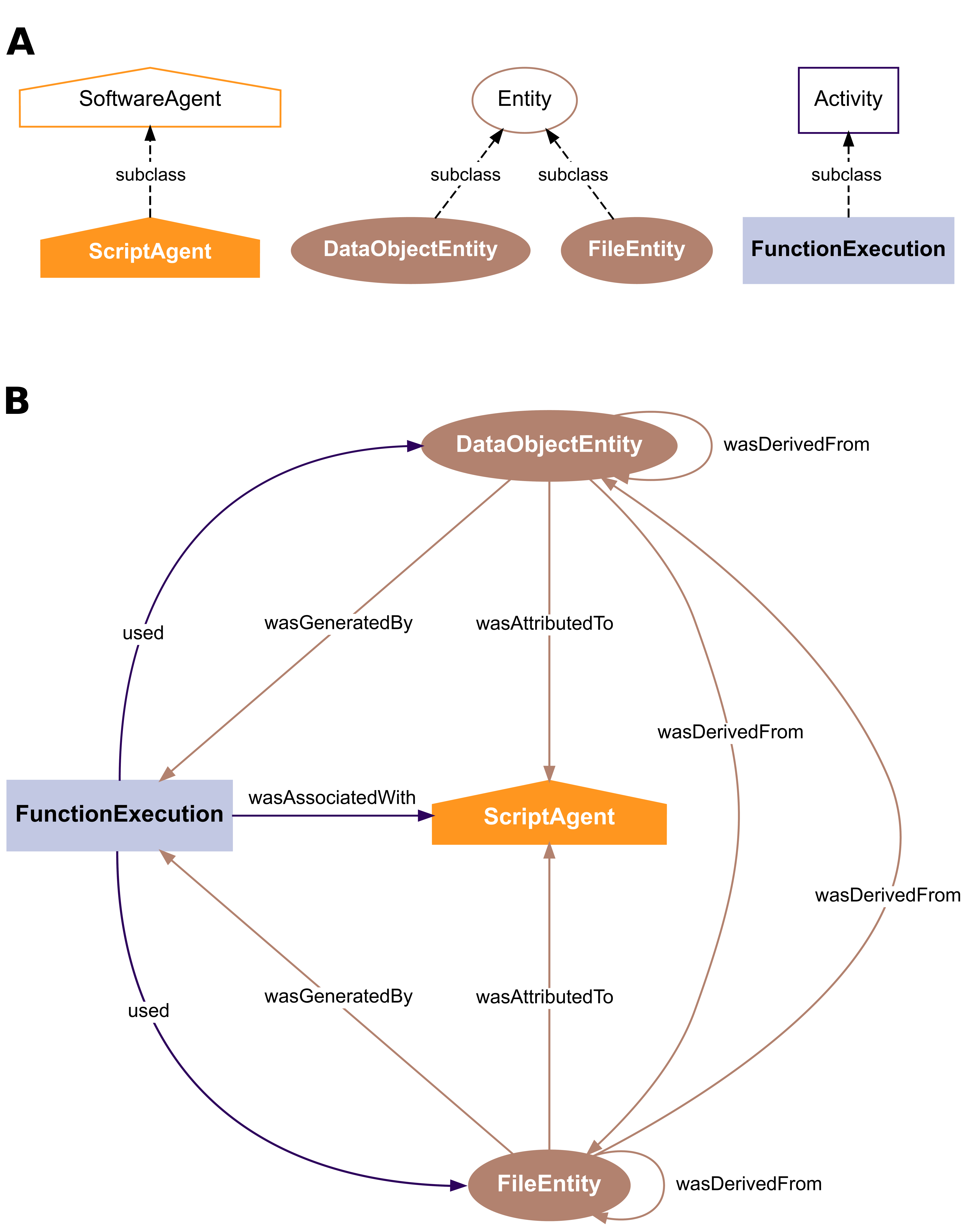}
    \caption{\textbf{The Alpaca ontology used to serialize provenance information. A.} Main classes (bottom, filled shapes) derived from PROV-O (top, unfilled shapes). Objects storing data and files are represented as PROV-O Entities. The execution of a function is a PROV-O Activity. The script is a PROV-O SoftwareAgent (which in turn is derived from the PROV-O Agent class). \textbf{B.} PROV-O provenance relationships among the classes in the Alpaca ontology.}
    \label{Figure 8}
    \end{fullwidth}
\end{figure}

For the RDF representation of the captured provenance, the PROV-O ontology \citep{W3C_PROV_O} was extended to incorporate properties relevant to the description of the provenance elements captured by Alpaca. \cref{Figure 8}A shows the main classes derived from the SoftwareAgent (a subclass of Agent), Entity, and Activity classes of the PROV-O ontology, and \cref{Figure 8}B shows the provenance relationships among the classes, as defined in PROV-O. These main classes are:

\begin{itemize}
	\item \alpacaDataObjectEntity: entity used to represent a \python object that was an input or output of a function;
	\item \alpacaFileEntity: entity used to represent a file that was an input or output of a function;
	\item \alpacaFunctionExecution: activity used to represent a single execution of one function with a set of parameters;
	\item \alpacaScriptAgent: agent used to represent the script that was run and executed several functions in sequence.
\end{itemize}

In addition to the classes derived from PROV-O, two additional classes are defined in the Alpaca ontology. They are used to represent specific information in the context of the provenance captured by Alpaca:

\begin{itemize}
	\item \textbf{Function:} represents a \python function. It contains code that is executed to perform some action in the script, and that can take inputs, parameters, and produce outputs (e.g., in our example, the \textit{welch\_psd} function defined in the \textit{spectral} module of the \elephant package);
	\item \textbf{NameValuePair:} represents information where a value is associated with a name. Name is a string and value can be any literal (e.g., integers, strings, decimal numbers). This is the main class used to store function parameters and data object metadata.
\end{itemize}

The Alpaca ontology also defines specific extended properties which are used to serialize function parameters, object/file metadata, and function information. They are summarized in \cref{table:alpaca_prov}.

\begin{table}
    \begin{fullwidth}
    \small
		\begin{tabular}{p{.2\linewidth} p{.2\linewidth} p{.3\linewidth} p{.2\linewidth}}
		\hline
		\textbf{Class} & \textbf{Property} & \textbf{Description} & \textbf{Value} \\
		\hline
		DataObjectEntity & hasAnnotation & the value of an annotation present in the object. Annotations are stored in dictionaries accessible by either the \textit{annotations} or \textit{array\_annotations} object attributes. The annotation name is the dictionary key, and the annotation value is the corresponding value. & NameValuePair \\
		\cline{2-4}
		& hasAttribute & the value of an attribute of the object (i.e., accessible by the dot such as \textit{signal.shape}) & NameValuePair \\
		\cline{2-4}
		& hashSource & one of the three methods used to obtain the object hash: \textit{joblib\_SHA1}, \textit{Python\_hash}, or \textit{UUID} & xsd:string \\
		\hline
		FileEntity  &  filePath & the absolute path where the file is located in the system & xsd:string \\
		\hline
		FunctionExecution & hasParameter & a parameter passed to the function when called & NameValuePair \\ 
		\cline{2-4}
		& executionOrder & value of the global execution counter when the function was executed & xsd:int \\
		\cline{2-4}
		& codeStatement & statement in the source code that originated the call to the function & xsd:string \\
		\cline{2-4}
		& usedFunction & function that was called & Function \\
		\hline
		ScriptAgent & scriptPath & absolute path to the file containing the source code of the script being executed & xsd:string \\
		\hline 
		Function & functionVersion & version of the package where the function is implemented. If function information is not available, it will be \textit{NA} & xsd:string \\
		\cline{2-4}
		& functionName & the name of the function, as written in the \textit{def} statement of the \python function definition & xsd:string \\
		\cline{2-4}
		& implementedIn & the full path to the module where the function is implemented (example: for the \textit{rand} function defined in the \textit{random} module of the \numpy package, the value of the property will be \textit{numpy.random}) & xsd:string \\ 
		\hline
		NameValuePair & pairName & name that identifies the value & xsd:string \\
		\cline{2-4}
		& pairValue & value that is associated with the name & rdfs:Literal \\
		\hline \ \\
		\end{tabular}
    \caption{\textbf{Properties of the classes defined by the Alpaca ontology.} The prefix \textit{xsd:} identifies the namespace of the XML Schema and \textit{rdfs:} the namespace of the RDF Schema. Values without a namespace indicated by a prefix are classes defined in the Alpaca ontology.}
    \label{table:alpaca_prov}
    \end{fullwidth}
\end{table}

For representing memberships, such as objects accessed from attributes (e.g., \textit{segment.analogsignals}), indexes (e.g., \textit{signals[0]}), or slices (e.g., \textit{signals[1:5]}), the PROV-O \textit{hasMember} property is used. The \alpacaDataObjectEntity representing the container object will have a \textit{hasMember} property whose value is the \alpacaDataObjectEntity representing the element accessed. The element will have one of the following properties to describe the membership:

\begin{itemize}
	\item \textbf{fromAttribute:} a string storing the name of the attribute used to access the object in the container (e.g., \textit{analogsignals} in \textit{segment.analogsignals});
	\item \textbf{containerIndex:} a string storing the index used to access the object in the container (e.g., \textit{0} in \textit{signals[0]}). This is not necessarily a number, as \python uses string indexes when accessing elements in dictionaries;
	\item \textbf{containerSlice:} a string storing the slice used to access the object (e.g., \textit{1:5} in \textit{signals[1:5]}).
\end{itemize}

In the RDF graph, each data object, file or function execution is identified by a Uniform Resource Name (URN) identifier \citep{RFC_8141}. The functions and script are also represented by their own URNs. To compose a unique identifier, specific information captured during the script execution is used in the composition of the final URN string. The \textit{authority} identifier element is a string that points to the institute or organisation which has responsibility over the analysis. It can be set using the \alpacaSettings function. The identifiers generated by Alpaca are summarized in \cref{table:identifiers}. 

\begin{table}
    \begin{fullwidth}
    \small
		\begin{tabular}{ p{.2\linewidth} p{.8\linewidth} }
		\multicolumn{2}{l}{A.} \\
		\hline
		\textbf{Alpaca ontology class} & \textbf{Identifier} \\
		\hline
		DataObjectEntity & urn:[authority]:alpaca:object:Python:[class name]:[object hash] \\ 
		FileEntity & urn:[authority]:alpaca:file:[hash type]:[file hash]  \\ 
		FunctionExecution & urn:[authority]:alpaca:function\_execution:Python:[script file hash]:[session ID]:[function name]\#[execution ID]  \\
		Function & urn:[authority]:alpaca:function:Python:[function name]  \\
		ScriptAgent & urn:[authority]:alpaca:script:Python:[script file name]:[script file hash]\#[session ID]  \\
		\hline{}
		\end{tabular}
		\bigskip \bigskip
		\begin{tabular}{ p{.2\linewidth} p{.8\linewidth} }
		\multicolumn{2}{l}{B.} \\
		\hline
		\textbf{Identifier element} & \textbf{Description} \\
		\hline
		authority & string defining the authority associated with the records \\
		class name & name of the object class in \python, with full module path from the source package where it is implemented \\
		object hash & content hash of the \python object \\
		hash type & method to hash the file (currently only SHA256 is supported) \\
		file hash & hash value of the file \\
		script file hash & SHA256 hash of the \python file containing the script source code \\
		session ID & UUID generated when activating Alpaca tracking (\alpacaSessionID) \\
		function name & name of the function, with full module path from the source package \\
		execution ID & UUID generated during the execution of the function (\alpacaExecutionID) \\
		script file name & name of the file containing the source code \\
		\hline{}
		\end{tabular}
    \caption{\textbf{Composition of URN identifiers for each element described in the Alpaca provenance records. A.} General schema for the composition of the identifier associated with each class in the ontology. \textbf{B.} Details of identifier parts mentioned between brackets in A.}
    \label{table:identifiers}
    \end{fullwidth}
\end{table}

\begin{figure}
    \begin{fullwidth}
    \centering
    \includegraphics[width=\linewidth]{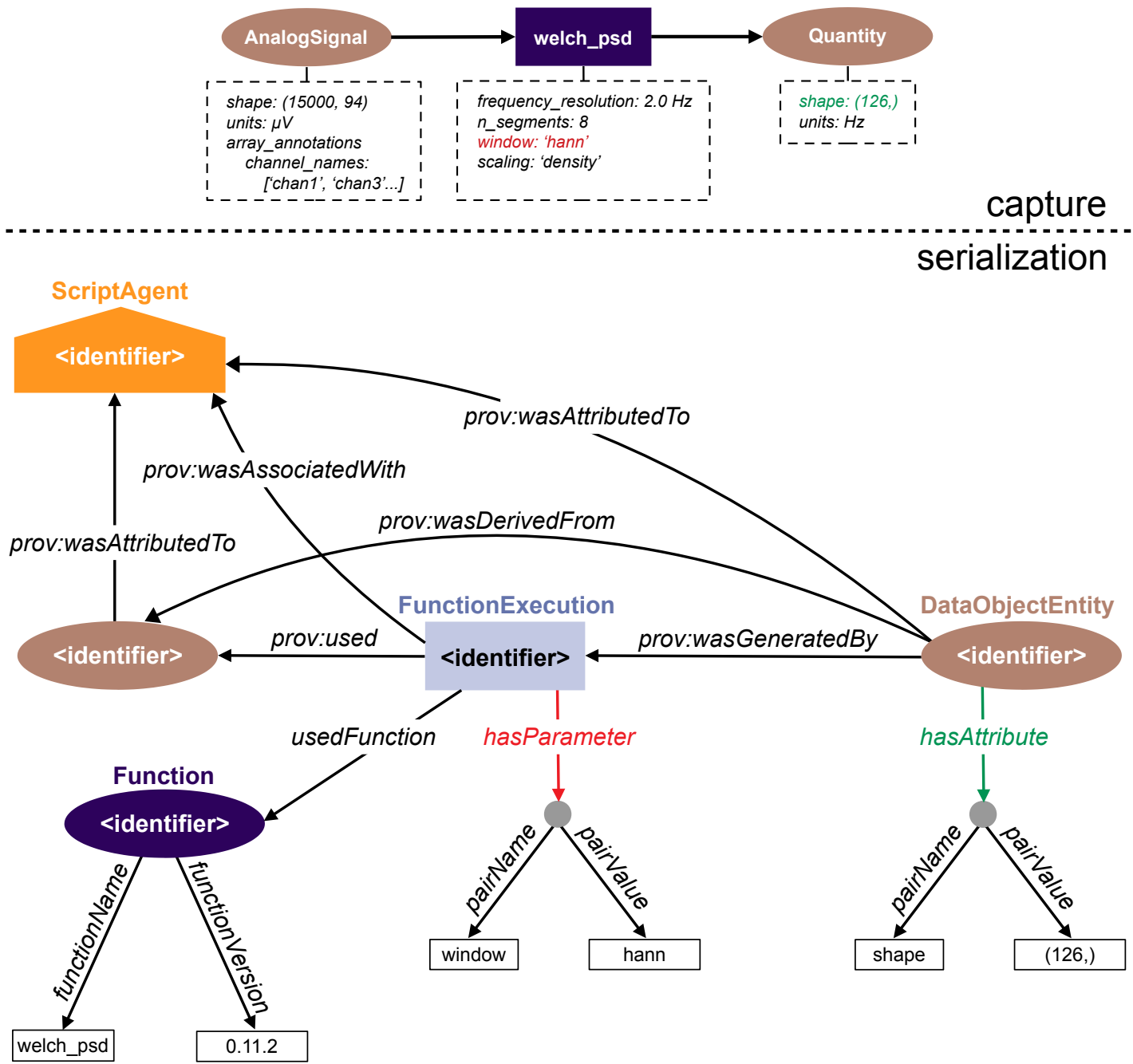}
    \caption{\textbf{Example of the serialization of a single function execution with the Alpaca ontology.} Top: Information captured during the execution of a function (\textit{welch\_psd}) taking a \neo \neoAnalogSignal as data input and returning a \quantity array. Bottom: The Alpaca ontology classes are used to represent the input/output objects and the function execution. The relationships from PROV-O are used to describe most relationships of the provenance. All elements are associated with the script as an agent. Object metadata and function parameters are serialized using the extended properties and relations provided by Alpaca (an example for a function parameter is shown in red using \alpacaHasParameter, and for an object attribute in green using \alpacaHasAttribute). In the diagram, the grey circles represent blank nodes of the \alpacaNameValuePair class. Some additional properties captured and serialized during the function execution were omitted in the diagram here for clarity. \textit{prov:} is the PROV-O namespace. Whenever a namespace is not defined, the class or property belongs to the Alpaca ontology.}
    \label{Figure 9}
    \end{fullwidth}
\end{figure}

\cref{Figure 9} summarizes how a single function execution is stored in the serialized RDF graph using the Alpaca ontology and the PROV-O properties.

\subsubsection{Visualization of the serialized provenance}

The provenance records serialized to RDF files can be loaded as \networkx\footnote{RRID:SCR\_016864; \url{https://networkx.org}} \citep{Hagberg08_11} graph objects. Besides the functionality for graph analysis offered by \networkx, the graph objects can be saved as GEXF\footnote{Graph Exchange XML Format, \url{https://gexf.net}} or GraphML\footnote{\url{http://graphml.graphdrawing.org}} files that can be visualized by available graph visualization tools, e.g., \gephi\footnote{RRID:SCR\_004293; \url{https://gephi.org}} \citep{Bastian09_361}, or other \python-based frameworks, e.g., \textit{Pyvis}\footnote{\url{https://pyvis.readthedocs.io}} \citep{Perrone20_4951}. This takes the advantage of existing free and open source solutions developed specifically for analyzing and interacting with graphs.

\begin{figure}
    \begin{fullwidth}
    \centering
    \includegraphics[width=\linewidth]{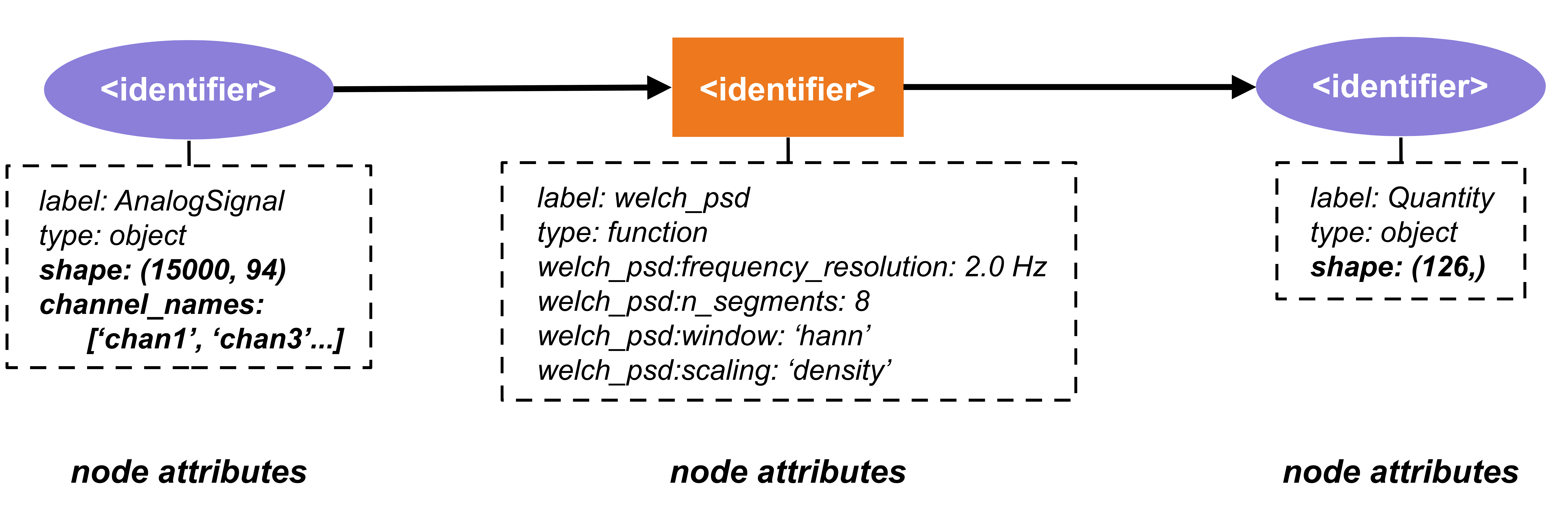}
    \caption{\textbf{Example of a \networkx graph generated from the RDF files serialized using the Alpaca ontology.} This shows the expected nodes (two \textbf{DataObjectEntities} and one \textbf{FunctionExecution}) from the example in \cref{Figure 9}. Node attributes of \textbf{DataObjectEntities} are included in the provenance trail if they are selected by the user when generating the graph using Alpaca functionality (attributes selected in the context of the use case scenario are shown in bold).}
    \label{Figure 10}
    \end{fullwidth}
\end{figure}

In Alpaca, the \alpacaProvenanceGraph class is responsible for generating the \networkx graph objects from serialized provenance data. \cref{Figure 10} summarizes how the visualization graph is obtained from the RDF graph. The resulting graph will have entities (\alpacaDataObjectEntity or \alpacaFileEntity) and activities (\alpacaFunctionExecution) as nodes, identified by the respective URN. Directed edges show the data flow across the functions. Metadata and function parameters are added to the attributes dictionary of each node. A few attributes are present for all the nodes in the graph (omitted in \cref{Figure 10} for clarity):

\begin{itemize}
	\item \textbf{type:} describes one of the three possible types of node: object, file, or function;
	\item \textbf{label:} for data objects, it is the \python class name (e.g., \neoAnalogSignal). For functions, it is the function name (e.g., \textit{welch\_psd}). For files, it is \textit{File};
	\item \textbf{Python\_name:} for data objects and functions, it is the full module path to the class or function, with respect to the package where it is implemented (e.g., \textit{neo.core.analogsignal.AnalogSignal}. For files, this attribute is not used;
	\item \textbf{Time Interval:} a string representing a time interval according to the standard used by \gephi that is composed from the order of the function execution. This information can be used to visualize the temporal evolution of the provenance graph, e.g., using the timeline feature of \gephi that displays only the nodes within a specified execution interval. 
\end{itemize}

The \alpacaProvenanceGraph provides options to tweak the visualization. First, it is possible to select which attributes and annotations from the metadata to include in the visualization graph. Second, parameter names can be prefixed by the function name, so that they can easily be identified. Third, nodes representing the builtin \python \textit{None} object (that is the default return value of a \python function) can be omitted. Finally, nodes describing a sequence of object access from containers (e.g., \textit{segment.analogsignals[0]}, which accesses the list in the \textit{analogsignals} attribute of \textit{segment}, followed by retrieving its first element) can be condensed such that a single edge describing the operation is generated. These visualization options reduce clutter and facilitate the visual inspection of the recorded provenance information. 

Finally, the provenance graphs can become large when repeated operations are performed within the script, such as using a \textit{for} loop to iterate over several data objects to perform computations. Therefore, an aggregation and summarization is available, adapted from the functionality already implemented in \networkx (from version 2.6). It uses the SNAP aggregation algorithm, and was modified from the original implementation to allow the selection of specific attributes of a set of nodes. Moreover, for functions executed with distinct set of parameters, the different values can also be taken into account when identifying similarity of nodes in summarizing the graph. The aggregation generates supernodes that represent not a single execution and data, but several identical or similar processing nodes. The identifiers of the individual elements that were aggregated in the supernode are listed in the \textit{members} node attribute. The total number of nodes aggregated into the supernode is stored in the \textit{member\_count} node attribute. In the end, the user can aggregate several nodes together, depending on whether they share the values of a given attribute, which allows the generation of a simplified version of the provenance trace that provides a more general overview of the analysis. 

\subsection{Code accessibility}

The code to reproduce the analyses presented as use case in this paper is freely available online at \url{https://github.com/INM-6/alpaca\_use\_case}. Figures 1, 2, 4-10 were manually created using Inkscape. Figures 3 and 13A are direct outputs of the corresponding scripts. Figures 11, 12 and 13B were created from graph visualization files generated by the corresponding scripts (GEXF format). The GEXF files were loaded into Gephi (version 0.9.7) and nodes were edited for color, position, and size. The graphs were exported to SVG files that were manually edited using Inkscape to compose the final figures. Editing involved adjusting label sizes and adding information available as node attributes in Gephi. The data used for the analysis can be found at \url{https://gin.g-node.org/INT/multielectrode\_grasp}.

\section{Results}

In the following, we will describe and evaluate the analysis provenance captured by Alpaca in the use case scenario described in \cref{sec:use-case}. After running \scriptfile with the code modified to use Alpaca, a detailed provenance trace was obtained and stored as \provfile. Corresponding GEXF graph files for visualization were generated, with distinct levels of aggregation and granularity of the steps in \scriptfile, ranging from a fine-grained view to a summarizing birds-eye view. The interactive analysis of those graphs using \gephi are presented in the form of a video (accessible at \url{\videourl}). Here, we will present the main features of the provenance trace using several \gephi graph exports. Then, we detail how they address the four challenges for tracking provenance of the analysis we identified in the Materials and Methods and \cref{Figure 1}.

\subsection{Overview of the captured provenance}

\begin{figure}
    \begin{fullwidth}
    \centering
    \includegraphics[width=\linewidth]{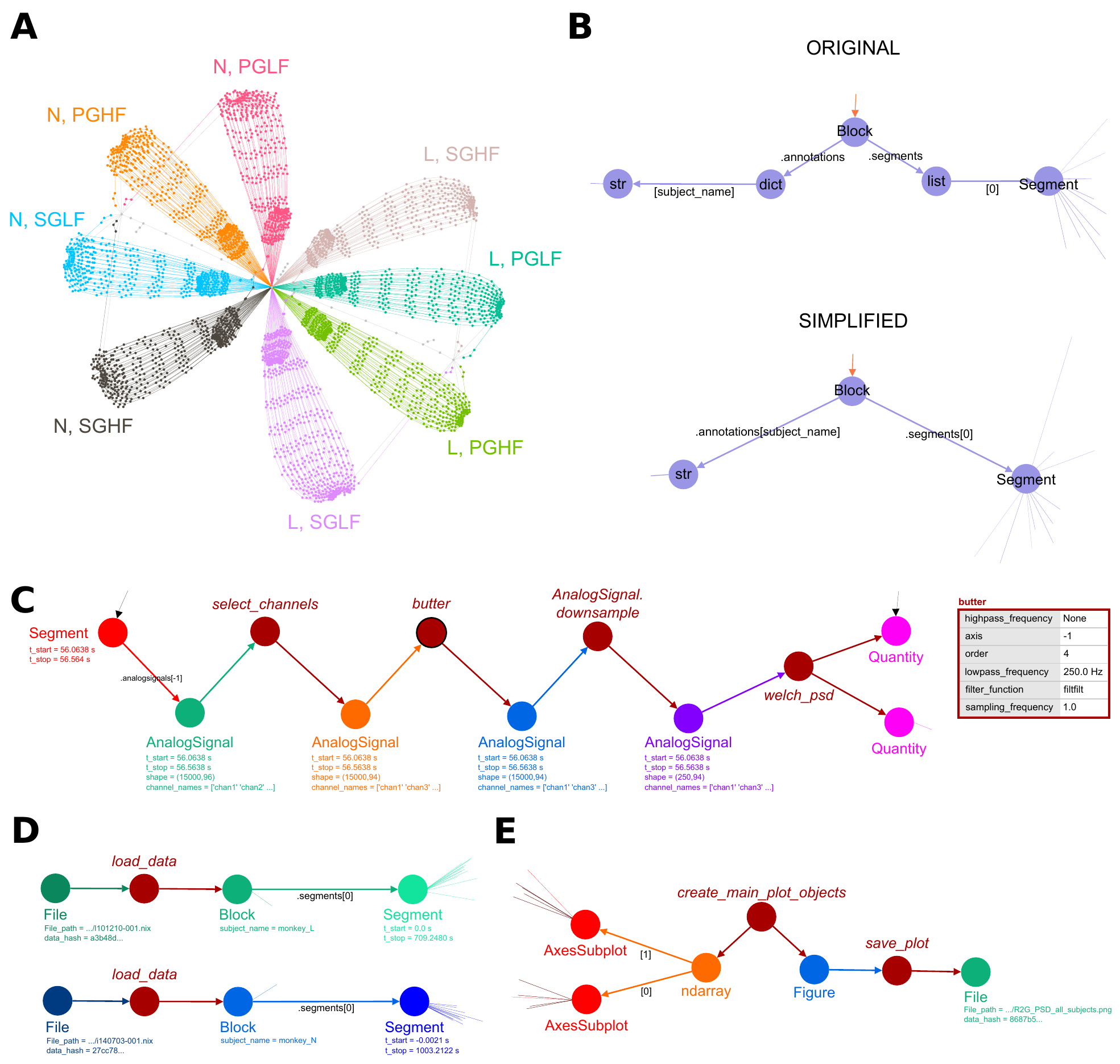}
    \caption{\textbf{Overview of the provenance captured by Alpaca stored in \provfile. A.} Visualization of the full (non-aggregated) graph corresponding to \provfile. Each graph region (color coded) corresponds to the processing for a single subject (monkey N or monkey L) and trial type (PGHF, PGLF, SGHF, or SGLF). \textbf{B.} Sequential object access operations are simplified such that  only a single edge represents attribute and subscript access. Original example (top): addressing the first \neoSegment object within the \neoBlock requires accessing its \textit{segments} attribute, followed by retrieving the first element (index 0). Simplified version (bottom): the edge shows the operation as \textit{.segments[0]}. The same simplification is applied to values stored in the annotations dictionary. \textbf{C.} Processing of an individual trial. Details of data objects and function parameters can be inspected. Values of selected attributes and annotations are shown below the data node labels. Function nodes are dark red with labels in italic. Exact parameters for the \elephantButter function execution are shown in the table. \textbf{D.} Loading of the \nix data file of each subject (monkey L: green; monkey N: blue). Function \textit{load\_data} (dark red) loaded each file and returned a \neoBlock object (\textit{subject\_name} annotation identifies the corresponding subject). The first \neoSegment of the \neoBlock was used for further processing. \textbf{E.} Generation of \plotfile from the \matplotlib \matplotlibFigure object. In D and E, the path and hash of files can be inspected.}
    \label{Figure 11}
    \end{fullwidth}
\end{figure}

\cref{Figure 11}A shows the overview of the graph generated from \provfile (\textit{None} objects returned by functions were removed). Overall, 3579 nodes and 4313 edges are present, and the graph has 8 colored regions. Each region corresponds to the iterations of the two outer loops in \scriptfile (i.e., loop over 2 subjects x loop over 4 trial types resulting in 8 iterations; \cref{Figure 4}). For the remainder of this study, the visualization is optimized to remove memberships due to the access of \neo objects in containers that introduces extra nodes in the graph. This simplification is illustrated in \cref{Figure 11}B.

Using the timeline feature of \gephi, it is possible to isolate specific parts of the graph (\cref{Figure 11}A) based on the execution order of statements in the \python code. Here, we single out the time window that corresponds to the processing of a single trial in a loop iteration (\cref{Figure 11}C) and then inspect individual attributes of the objects and parameters of the functions involved until the computation of the PSD. It is possible to inspect the start and end time points of the trial segment with respect to the recording time in the dataset using the \textit{t\_start} and \textit{t\_stop} attributes of the \neoSegment object at the beginning of the trace, thus uniquely identifying the analyzed data segment. It is also possible to review the \neoAnalogSignal object containing the data that was later processed and used to compute the PSD by the \elephantWelch function of \elephant. General attributes, such as the shape of the data array of the \neoAnalogSignal object, can be accessed together with specific metadata, such as the names of the channels associated with the time series in the data. Finally, for these intermediate steps, it is possible to inspect specific parameters passed to each function: the attributes of \alpacaFunctionExecution graph nodes (shown example: \elephantButter) corresponding to function parameters are prefixed by the function name, followed by the name of the argument as defined in the \python function definition (cf., \cref{Figure 10}). Taken together, Alpaca captured these types of information for each individual step throughout the execution of \scriptfile such that each iteration of the central analysis can be traced in detail after completion of the script.

\begin{figure}
    \begin{fullwidth}
    \centering
    \includegraphics[width=\linewidth]{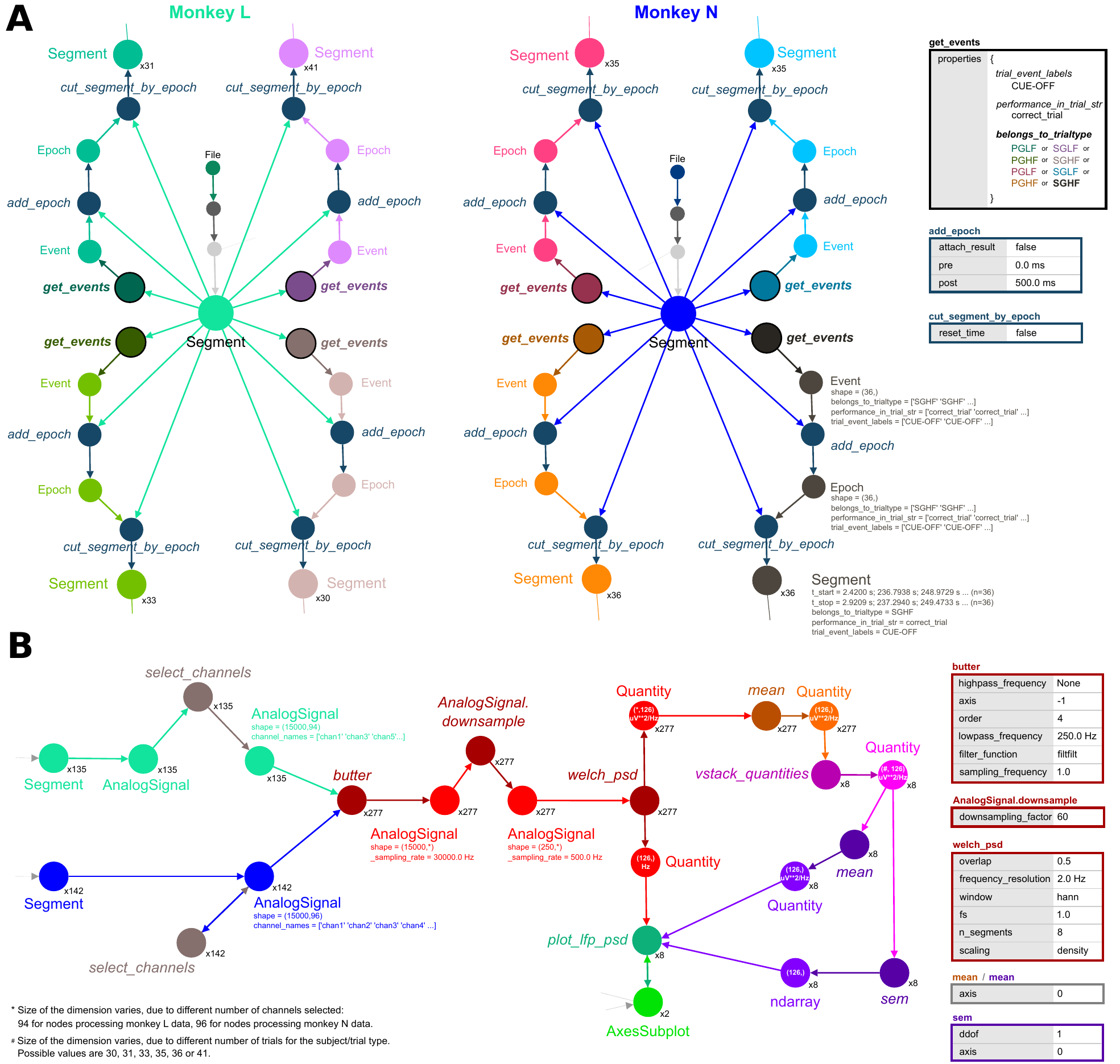}
    \caption{\textbf{Alpaca captured the detailed provenance of \plotfile.} Function nodes are labeled in italic and data objects in normal style. The multipliers next to a node show how many similar nodes were aggregated at that step. Selected function parameters are shown in the tables. \textbf{A.} The aggregated graph demonstrates common steps in preprocessing the \neoSegment objects containing data of monkey L (left) and monkey N (right). For each trial type, the \neo function \textit{get\_events} extracted the times of the CUE-OFF events of correct trials (see \textit{properties} parameter in the table; dictionary key names are in italic). Each execution of \textit{get\_events} used distinct \textit{belongs\_to\_trialtype} values (color in table matches color in graph). \neo functions \textit{add\_epoch} and \textit{cut\_segment\_by\_epoch} were called with same parameters across monkeys/trial types to extract data as \neoSegment objects corresponding to 500~ms after CUE-OFF (dark blue). The number of trials can be inferred from the multipliers of the \neoSegment objects. The details for the preprocessing of monkey N/SGHF trials are shown (dark grey). \textbf{B.} Analysis steps after cutting data into trials (aggregated for each monkey/trial type). The values of the \textit{shape} and \textit{units} attributes of \quantity objects are shown inside the node. Function \textit{select\_channels} was executed taking the \neoAnalogSignal with the electrode data for the trial (dark brown; n=135 trials for monkey L and n=142 trials for monkey N). (continues on next page)}
    \label{Figure 12}
    \end{fullwidth}
\end{figure}

\begin{figure}
    \begin{fullwidth}
    \contcaption{(continued) For monkey L, channels 2 and 4 (\textit{'chan 2'} and \textit{'chan 4'} values in the \textit{channel\_names} array annotation) were removed. All trials (n=277) were processed similarly (red shades), consisting of applying a 250~Hz low-pass Butterworth filter (function \elephantButter), downsampling to 500~Hz, and computing a PSD using the Welch method using a Hanning window with 50\% overlap and 2~Hz frequency resolution (function \elephantWelch). For each \quantity array with power spectra, an average across channels was obtained (orange shades) and for each monkey/trial type (n=8) those estimates were combined to obtain an array for that trial type (pink shades; the shapes of the arrays match the number of trials obtained in the pre-processing). Mean and SEM across trials were obtained from these arrays (purple shades) and plotted (green shades).}
    \end{fullwidth}
\end{figure}

It is possible to retrace the first steps after loading the two data files (\cref{Figure 11}D). A function called \textit{load\_data} (defined in \scriptfile) was called with the neural data file (available as a dataset in the NIX file format) of one particular subject as input and returning a \neoBlock object with all the data of that recording session. We can inspect the \textit{subject\_name} annotation of \neoBlock and identify in human readable text which subject corresponds to each object. We can alternatively bind each \neoBlock to the specific source data file, by inspecting the \textit{File} node associated with each object, and obtain the SHA256 hash (\textit{data\_hash} node attribute). Although the actual path used in the analysis (\textit{File\_path} node attribute) will point to the actual location of the file in the system where the script was run, the hash will allow the identification of the file regardless of its name and location. Moreover, the graph shows that the first \neoSegment stored in the \neoBlock was accessed (through the \textit{segments} attribute of \neoBlock), and this was the main source for all subsequent analysis done for each monkey. By inspecting the node of the \neoSegment object, we have access to its attributes and annotations, such as the starting and end times of the data in recording time (-0.0021 and 1003.2122 seconds for subject monkey N, and 0.0 and 709.2480 seconds for monkey L; \cref{Figure 11}D).

In a similar way as described above for reading the input data, we can inspect the generation of the output file (\cref{Figure 11}E). It was obtained from a \matplotlib \textit{Figure} object that was initialized by a function at the beginning of the script execution (\textit{create\_main\_plot\_objects}), was successively filled with graphs as power spectra were calculated, and finally saved to disk as a PNG file using a function called \textit{save\_plot}.

\subsection{Understanding the data preprocessing}

\cref{Figure 12}A shows the sequence of steps applied to the \neoSegment object that contains the full data for one subject. When aggregated by function parameters (i.e., simplified based on similarity of function parameters), the graph shows four separate paths that start from each of the two \neoSegment objects (one per subject). Each is comprised of the \neo functions \textit{get\_event}, \textit{add\_epoch}, and \textit{cut\_segment\_by\_epoch}. Each of those functions performs a specific action: identify specific events during the recording (stored in an \neoEvent object) according to selection criteria, select a window of data around these identified event timestamps (stored in an \neoEpoch object), and finally use the windows stored in the \neoEpoch object to cut the large \neoSegment, producing one \neoSegment object per epoch containing a window.

We can now analyze the captured provenance to verify the detailed parameters used in each of those preprocessing functions. \textit{get\_event} used a parameter called \textit{properties}, together with the \neoSegment object as input. That parameter defines a dictionary with keys and values that are compared to the annotations or attributes of a \neo \neoEvent object in order to select the desired subset of all events recorded during the experiment. All four paths considered the CUE-OFF event of correct trials (defined by the \textit{trial\_event\_labels='CUE-OFF'} and \textit{performance\_in\_trial\_str='correct\_trial'} dictionary entries). However, in each path, the function was called with the \textit{belongs\_to\_trialtype} value containing one of the four possible trial labels: PGHF, PGLF, SGHF, or SGLF. Therefore, each \neoEvent object returned by \textit{get\_event} will contain the times of CUE-OFF of all correct trials of one of the four trial types.

The times of the generated \neoEvent objects were used to define epochs and cut the data to obtain segments of the trials of a particular type. Inspecting the subsequent executions of the functions \textit{add\_epoch} and \textit{cut\_segment\_by\_epoch}, their parameters show that epochs were defined as 500~ms after the CUE-OFF event (\textit{pre=0.0 ms} and \textit{post=500.0 ms}), and the absolute recording times were preserved when cutting (\textit{reset\_time=False}). Therefore, for each subject, we can partition the provenance graph in four separate paths, each dealing with processing data of a particular trial type (the outer loops of \scriptfile; \cref{Figure 2}). Not only these selection criteria for extracting the data are retained by the provenance trail, but also we can retrieve the precise time points used for cutting the data and calculated only during run-time based on the loaded data on a trial-by-trial basis (by inspecting the \textit{t\_start} and \textit{t\_stop} attributes of each \neoSegment generated by \textit{cut\_segment\_by\_epoch}). Overall, Alpaca allowed us to understand the initial data preprocessing and trial definitions, addressing challenges 1 and 2.

\subsection{Inspecting the data flow used to generate a result}

The figure stored in \plotfile and shown in \cref{Figure 3} could have been produced by different versions of \scriptfile, with steps in different order or new steps added. A likely scenario is the necessity to filter out some channels for one of the datasets. In \cref{Figure 12}B, we see that for each subject, a user-defined function called \textit{select\_channels} was applied to the data. For monkey L, it is apparent from the shapes of the data arrays that 2 recording channels were excluded (due to signal quality), such that only 94 of the 96 recording channels were used. The provenance track captured by Alpaca shows this, as the returned \neoAnalogSignal object is different from the object containing all the channels, and the \textit{shape} attribute shows the removal of the two channels.

At this point, it is possible to bind \provfile to \plotfile through the SHA256 hash of the file written by the function \textit{save\_plot} (\cref{Figure 11}E). \provfile will also have all the function executions linked to the script identifier, obtained from the hash of \scriptfile and \alpacaSessionID (cf. \cref{table:identifiers}). Thus, it was possible to record all operations within a single script together with the actual parameters used. In this way, the provenance information can be used to automatically capture and retain the ongoing development process from the perspective of the generated results, addressing challenges 2 and 3.

\subsection{Reviewing analysis parameters}

In between runs of a single version of \scriptfile, the analysis parameters could also have been changed and leading to alternate versions of the PSD estimates in \plotfile generated by each run. A scenario where this is likely to occur is one where the scientist performing the analysis may have interactively iterated the code several times to find a set of parameters that allowed a good visualization of the power spectra.

The provenance track captured by Alpaca allows to inspect the values of each individual function call. From the \neoAnalogSignal object after channel selection, there is a common pathway in the aggregated graph for both subjects (\cref{Figure 12}B). The functions \textit{butter}, \textit{AnalogSignal.downsample}, and finally \textit{welch\_psd} were called sequentially. Those correspond to the filtering, downsampling, and computation of the PSD using the Welch method. Each of those functions have key parameters that will affect the PSD estimate, and the parameters were captured automatically.

We can use the provenance information to verify that a 250~Hz low pass cut-off was used for the filtering (from the parameter passed to the \elephant \elephantButter function). Moreover, we verify that the signal was downsampled by a factor of 60 (method \textit{downsample} from the \neoAnalogSignal object). By inspecting the shapes of the \neoAnalogSignal objects that are input and output of the function, we can verify the downsample operation: the input object had 15000 samples and, after \textit{AnalogSignal.downsample}, the number was reduced to 250.

Finally, it is possible to inspect all the parameters for the PSD computation using the \elephant \elephantWelch function: a Hanning window was used, for an estimate with a 2~Hz frequency resolution. The resulting objects storing the frequency bins and power estimates (\quantity arrays) are discernible by the \textit{units} attribute. The frequency array has a dimension of 126, which is expected for a PSD of a continuous signal downsampled to 500~Hz and with a frequency resolution of 2~Hz. It is also possible to observe that the power estimates are a two-dimensional array with first dimensions of 96 (for monkey N) and 94 (for monkey L), which agree with the source \neoAnalogSignal objects and indicate the number of channels. Therefore, the power estimates were obtained for each channel as a single array.

Addressing challenges 1 and 2, it possible to retrieve the value of any parameter that may have resulted from trial and error iterations during the development of \scriptfile, as the provenance information shows the detailed history of the generation of the data objects that were ultimately used by the plotting function.

\subsection{Facilitating sharing of analysis results}

When sharing \plotfile with others, some parts of the figure therefore leave guesswork to the collaborator. However, \provfile contains several pieces of information that are not accessible from the figure stored in \plotfile alone. 

In addition to the details of the analysis steps presented above, it is also possible to know the last steps used to transform the data before plotting the lines and intervals using the \scriptPlotLfpPsd function (\cref{Figure 12}B). First, an average of the power across all channels was obtained for each trial. The \numpy \textit{mean} function was applied to the array with the per-channel power estimates, over the first axis (\textit{axis=0} parameter). Then, the averages of all trials of the same trial type of a single subject were averaged in a grand mean (using the \numpy \textit{mean} function). The individual trial averages were also used to obtain a SEM estimate (using the \scipy \textit{sem} function). Finally, the grand mean and SEM were passed to the \scriptPlotLfpPsd function that performed the plotting in the \matplotlibAxes object corresponding to the graph panel for that subject, taking the multiplier 1.96 as a parameter to define the width of the intervals. Not only all these steps are now apparent, but it is also possible to know how many trials were used for each subject when plotting (monkey N: PGHF=36, PGLF=35, SGHF=36, or SGLF=35; monkey L: PGHF=33, PGLF=31, SGHF=30, or SGLF=41; \cref{Figure 12}A and \cref{Figure 12}B). In addition, for each call of \scriptPlotLfpPsd it is possible to inspect the parameter providing the legend label with respect to the source of the mean, SEM and frequency data used as inputs.

As mentioned above, two electrode channels were excluded in the analysis of monkey L data. The provenance information in \provfile makes it possible to check the \textit{channel\_names} annotations of each \neoAnalogSignal object used in each iteration when computing the PSD (\cref{Figure 12}B). The inspected labels show that channels 2 and 4 were excluded for this monkey.

\begin{figure}
    \begin{fullwidth}
    \centering
    \includegraphics[width=0.95\linewidth]{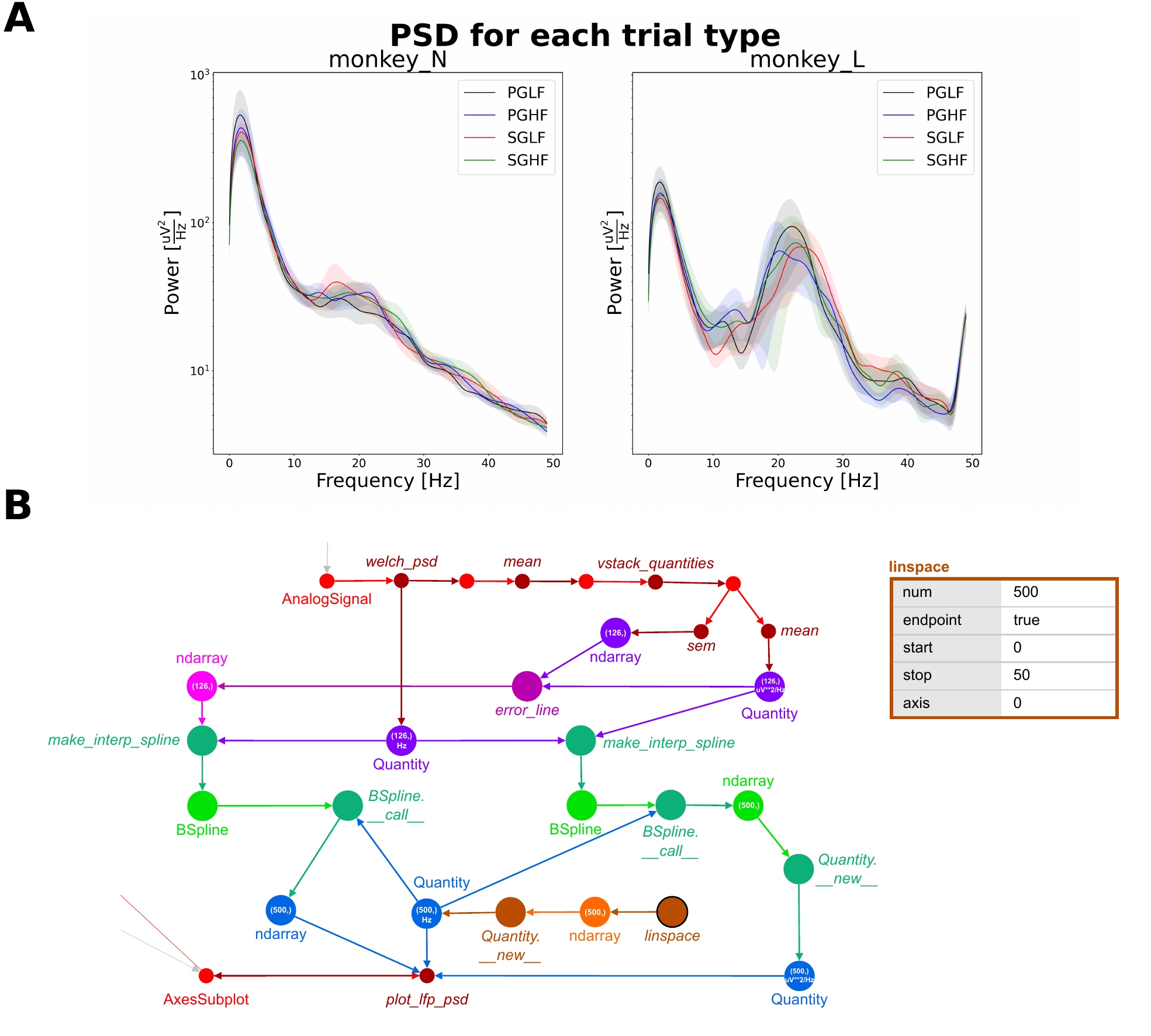}
    \caption{\textbf{Provenance captured by Alpaca shows transformations of the results before plotting. A.} Alternate version of \plotfile. The lines appear smoothed in comparison to the plot shown in \cref{Figure 3}. \textbf{B.} Provenance track showing the steps after the computation and aggregation of the PSDs across channels and trials, before plotting with \scriptPlotLfpPsd. The visualization graph was aggregated to demonstrate the common steps when plotting the different estimates for each subject and trial type. Nodes in darker shades with labels in italic are function calls and nodes in lighter shades with labels in normal style are data objects. The steps also present in the generation of the plot in \cref{Figure 3} are highlighted in red (top; some labels omitted for clarity). The \numpy array (with SEM estimates) and \quantity arrays (with PSDs or frequencies) used in the original \plotfile are shown in purple (array shape and physical quantity are shown inside the nodes). In contrast to the generation of the original plot, an additional function computed the error line values (pink shades). Next, the \numpy arrays with error estimates or the \quantity arrays with PSD estimates were passed together with the \quantity array with frequencies to the \scipy function \textit{make\_interp\_spline}, which generates a \textit{BSpline} interpolation object (green shades show the interpolation steps). For the interpolation using \textit{BSpline}, an array with 500 elements between 0 and 50 was generated with \numpy \textit{linspace} function (orange shades; the function node is highlighted with a black border). The detailed parameters for \textit{linspace} recorded by Alpaca are shown in the table. This \numpy array was converted to a \quantity array that was used with the interpolation objects to obtain the final error and PSD arrays used in the plots (blue nodes).}
    \label{Figure 13}
    \end{fullwidth}
\end{figure}

An additional scenario to illustrate how to make use of the captured provenance in a shared environment is presented in \cref{Figure 13}A. Here, a plot resembling the one presented in \cref{Figure 3} is stored in \plotfile. However, the lines and interval area boundaries appear smoothed, suggesting the plot was generated by an alternate version of \scriptfile. The provenance captured by Alpaca reveals steps after the aggregation of the power estimates across trials. Spline smoothing objects from the \scipy package were used to generate new arrays that were the inputs to the plotting function \scriptPlotLfpPsd (\cref{Figure 13}B). With this information, collaborators receiving \plotfile can clearly identify that the plot is not showing the actual estimates but a smoothed version.

In summary, addressing challenge 4, the provenance information captured by Alpaca facilitates sharing \plotfile as it provides additional information for finding and understanding the results without requiring extra work by the scientist performing the analysis.

\section{Discussion}

In this paper, we presented Alpaca, a toolbox to capture fine-grained provenance information when executing \python code, with a specific focus on scripts that analyze data. The information is saved as a metadata file that represents a sidecar file to the saved analysis results. Using a realistic use case analysis of calculating power spectra estimates in a massively parallel electrophysiology dataset, we showed how this captured provenance metadata helps in understanding an electrophysiology analysis result that could ultimately be shared among collaborators. With the help of graph visualizations, it is possible to inspect the data flow across functions together with other details that were available at run time, such as object attributes and annotations and function parameters. The toolbox takes advantage of existing standards to represent electrophysiology data in \python (e.g. \neo) by also capturing relevant object metadata into the provenance records. In the end, it was possible to obtain detailed information that were not available from the result file alone. This provided a better context for the interpretation of an analysis result and adds to the rigor in its reuse.

In the beginning, we introduced four challenges associated with the analysis of electrophysiology datasets that we aimed to consider in designing a toolbox to capture provenance. We then showed, using our concrete use case, that Alpaca addresses these challenges. First, the customized data preprocessing routine using functionality of the \neo package was described in the provenance record with all the relevant parameters. Second, any state of the parameters of the functions called in the script and the data flow will be automatically recorded together with the results to allow detailed comparisons as the script is developed and adapted over time. Third, in agile, interactive analysis scenarios changes to the source code or execution order of code blocks leading to different result files and different provenance tracks that can be bound to the result files and code by the file and script identifiers. Finally, Alpaca provided a structured provenance record describing the history of generation of \plotfile as an additional file that is suitable for sharing together with the results. This serialized provenance makes not only information available in the plot in \cref{Figure 3} (e.g., subject names, units), but also that were not apparent at all (e.g., the annotations employed to select the timestamps of the CUE-OFF events that are the start time of the trial data used) accessible in a machine-readable format that can be inspected by scientists receiving the shared analysis results. Overall, the provenance information captured by Alpaca delivers the information required for understanding an electrophysiology analysis result, facilitating especially work in collaborative environments.

Trust is a key factor in experimental data analysis, especially in collaborative contexts. Result artefacts (files, figures, \ldots) are useful as long as the processes that generated them fit the hypotheses and research questions that guided the analysis in the first place. As provenance information describes the data and its transformations, it is expected that it should help in building trust in the analysis of electrophysiology data. The provenance information captured as a metadata file by Alpaca helps in that direction. With the example presented in this paper, we demonstrated that the toolbox describes the analysis processes in detail, reducing uncertainty on every step of the data analysis. Data loading, preprocessing, signal processing, obtaining the actual PSD estimates, and preparing the data for plotting and ultimately saving the result file were apparent when analyzing the provenance records saved as \provfile. In addition, the key parameters that determine each intermediate result are clearly defined. In the end, Alpaca contributes to building trust in the processing of analyzing data in collaborative environments and sharing results among peers.

Alpaca might improve the reproducibility of the results when analyzing electrophysiology data. Considering reproducibility as the ability to reproduce a given analysis result by different individuals in different settings, the detailed information provided by Alpaca provides a good description of the processes involved in the generation of the analysis result even in the absence of the original script. Although a full re-execution or reconstruction of the source code is neither possible nor the goal of the tool, still it is possible to know the sequence of functions used, their source packages and versions, and the relevant parameters in a level of detail that would help in any reimplementation of the analysis pipeline from scratch. The provided identifiers and hashes would also help in checking whether the data objects are equivalent between runs, without having to serialize the full object data at each step. In the end, although the generation of the exact result file will require the re-execution of the original script, the information summarized by Alpaca already makes any attempts to reproduce the results using a different code more likely to succeed.

Alpaca also contributes to make the electrophysiology data analysis results more compliant to the FAIR principles \citep{Wilkinson16_1}. These were developed to provide recommendations and requisites to increase the findability, acessibility, interoperability and reusability of data. While typically considered in the context of the source data files obtained from an experiment, the principles could be extended to include artefacts such as a result stored in \plotfile. Indeed, increasing the FAIRness of such electrophysiology analysis results would bring several benefits. First, if the results are findable, it is easier to navigate among a collection of results such as hundreds of files in a shared folder. Second, the interoperability would allow for the comparison of similar results produced by different implementations of a single method (such as the case of different \python toolboxes providing similar analysis functions, such as the computation of a PSD using the Welch method that is available in \elephant, \scipy, \textit{MNE} \citep{Gramfort13_1}, and many others). Finally, the reusability of the results would eliminate the necessity of repeating required analyses when they were already performed. This could be the case of the use case presented in this paper, where a collaborator might be interested in using the PSD estimates as a starting point for further analyses of the same experimental datasets. If the existing \plotfile already provided an adequate analysis with respect to the preprocessed trial data, signal processing, parameters of the PSD estimates and aggregation over channels and trials, she could simply reuse it to make any required inferences before starting her analysis. Alpaca provides advances mainly with respect to the reusability FAIR principle, as the analysis results are obtained with detailed provenance, and the results are also described with accurate and relevant attributes such as the annotations present in the \neo data objects. However, Alpaca also improves the interoperability and findability of the results. Regarding interoperability, first the provenance information is structured in a machine-readable format, using the PROV provenance model that defines a broadly-used vocabulary for provenance representation. Moreover, the metadata (in the form of attributes and annotations of the data objects) and function parameters (that can be seen as a special kind of metadata when considering what is proposed in the FAIR principles) are also structured in a machine-readable format defined formally in the Alpaca ontology. Finally, the findability of the results is improved, as Alpaca binds the identifiers of the individual data objects, files, script, functions and function executions to the analysis outcome, making it queriable via, e.g., the functions used in generating the outcome or by specific parameter settings. In the end, although a fully FAIR-compliant solution requires the development of additional resources such as controlled vocabularies and ontologies to represent the electrophysiology data analysis processes, Alpaca already provided increased adherence of the electrophysiology analysis result to the FAIR principles. 

Besides those improvements associated with the machine-readability of the captured provenance, Alpaca also facilitates the access to the provenance of the analysis results by humans. The visualization graphs generated from the RDF files eliminate the necessity of complicated tools such as SPARQL queries to extract and interact with the captured provenance. This ability to explore the provenance graphs and inspect data object attributes and annotations as well as function parameters allows the scientist to visually understand the details of each individual data transformation which facilitates the interpretation and understanding of the analysis result. This is complemented by the possibility to aggregate similar nodes in the graphs producing summarizations. While these lose the fine-grained details, they provide a high-level overview that is more descriptive of the analysis process than any accompanying textual documentation or the script source code. Ultimately, Alpaca not only records the provenance information for documentation purposes, but it is also a powerful tool to understand the history of the analysis result.

One design feature of Alpaca is that it does not provide a description of the control flow in the script. This is apparent from the main structure of the provenance graph of the example presented in \cref{Figure 11}A, where each iteration of a \textit{for} loop appeared as a separate path starting from the function that generated the objects accessed in the loop. From the implementation perspective, the same graph would be obtained if the source code was structured in a way that the access of individual elements was done without a loop (i.e., instead of looping over a container with N elements, insert N function calls, each using a different element from the container). Therefore, at this point, it is not possible to use the saved provenance to make inferences about the code. In contrast, the data-centric approach taken by Alpaca was developed with the aim of exposing the data and its transformations, and relevant parameters and metadata. Thus, we consider that the resulting provenance lacks complexity while making the data flow clear, regardless of the control flow used to achieve it.

\subsection{Comparison with existing tools}

There are existing tools that aim to capture and describe provenance during the execution of scripts, and each tool has distinct technical approaches and aims to accomplish distinct objectives (see \cite{Pimentel19_47:1} for a review). One approach is to capture provenance during the script run time, as adopted by Alpaca. In this context, we highlight \noworkflow \citep{Murta15_71}, as it was intended to be used in a similar scenario than Alpaca, i.e., the execution of standalone \python scripts that analyze data and produce output files. However, in contrast to Alpaca, \noworkflow does not require code instrumentation, but relies on a custom command line tool to run the script. The \noworkflow tool performs an \textit{a priori} analysis of the code together with tracing during the script execution to provide a very in-depth description of the sequence of functions called and to generate a detailed call graph as provenance information. All the information is captured and saved in a local database. The focus of \noworkflow is storing and describing repeated runs of the code (trials), highlighting the differences and evolution across trials. Although \noworkflow provides a very detailed description of the analysis process at the level of every function call (which is not possible for Alpaca as it tracks only the functions identified by the decorator), it falls short for some aspects introduced by Alpaca. First, we decided to save provenance using a data model derived from PROV, which increases interoperability, while \noworkflow currently relies on a custom relational database to structure the information on the function executions. Moreover, Alpaca aims to provide an extended description of the data objects across the script execution, which was implemented in the ontology used in the RDF serialization. Together with the description of the sequence of functions executed, this additional information is relevant for the understanding of the analysis result, especially regarding metadata provided as annotations. An example in the presented use case is the identification of the data pertaining to the individual trial types. \noworkflow would have shown the loops and sequence of \neo functions used to cut the data into the smaller trial segments, but the annotations identifying each \neoEvent object used for the preprocessing using those functions would not be accessible. In the end, this relevant information is accessible from the provenance records provided by Alpaca. Overall, Alpaca captures provenance with a different perspective on the analysis process, that is more relevant for the particularities of electrophysiology data analysis as introduced at the beginning of this paper. 

\aiida \citep{Pizzi16_218a} is another tool that can be used to capture provenance in data analysis workflows implemented in \python. It was developed as a complete solution for the automation, management, persistence, sharing and reproducibility of complex workflows. With respect to data provenance, \aiida tracks and records the inputs, outputs, and metadata of computations and produces a complete provenance graph. The technical approach is similar to Alpaca since it also uses decorators to instrument the code. However, \aiida has other design features: (i) it saves provenance in a centralized storage; (ii) as part of the provenance tracking, any data object can be saved to the database with a unique identifier, allowing its retrieval later for reuse together with the lineage. In the end, \aiida is a more holistic tool for reproducibility than Alpaca, as it is possible to re-execute the analysis using the same data objects previously stored. However, we also identify limitations in comparison to Alpaca. First, \aiida requires any existing data objects (such as the ones provided by the \neo framework) to be wrapped by custom objects so that the system can identify and serialize their content to the database, which can be achieved through a plugin system. This means that the user must implement this interface for any and every specific data object in a custom framework. Not only this requires a considerable amount of effort but this may also introduces a level of maintenance complexity as the data framework evolves and the user needs to ensure that the wrappers retain compatibility in the future. With the approach taken by Alpaca, we tried to keep the original \python objects without any fundamental transformation in their structure, and therefore we focused on identifying them using the URNs so that the lineage graph can be constructed, together with the description of their relevant metadata. An additional limitation of \aiida is the overall setup of the system to obtain the provenance information. In the approach taken by Alpaca, the provenance information is saved locally as RDF in an additional file that should accompany the actual results produced by the script, using the interoperable PROV data model. On one hand, one may argue that this is annoying as any sharing requires the user to also share the provenance metadata file together, which is less convenient than just querying a database using a command line tool such as the one provided by \aiida. On the other hand, this adds simplicity to use the tool as no special services are required to be set up at the user system. It is important to note that, at this point, the individual RDF files produced by Alpaca could also be stored into a centralized RDF triple store system (either locally or remote) in order to provide similar functionality, if desired. Finally, a third limitation is the use of a non-interoperable standard for description of provenance, as the provenance graphs by \aiida rely on a custom description of the data and control flows, and obtaining the provenance graphs requires the user to query the information using the specific \aiida API as opposed as using a standard such as SPARQL. In the end, in comparison to \aiida, Alpaca has a reduced entry barrier to implement provenance tracking into existing scripts, which may be relevant for the average electrophysiology lab to start benefiting from provenance capture during the analysis of their experimental data. It is likely that each of the two tools focus on the needs brought by different application scenarios, such as a small lab versus a large research institute. For the small lab, improvements in collaborative work in the analysis of electrophysiology data by capturing more detailed provenance might be quickly achieved by using a tool like Alpaca.

Recently, \caesar (CollAborative Environment for Scientific Analysis with Reproducibility) was proposed as a solution for the end-to-end description of provenance in scientific experiments \citep{Samuel22_e921}. The overarching goal of \caesar is to capture, query and visualize the complete path of a scientific experiment, from the design to the results, while providing interoperability. This was achieved by the implementation of the REPRODUCE-ME model for provenance \citep{Samuel22_1}, based on existing ontologies such as PROV-O \citep{W3C_PROV_O} and P-Plan \citep{Garijo12}. A solution called \ProvBook is also provided in order to support reproducibility and to describe the provenance of the analysis part of the experiment implemented as \jupyter notebooks. Alpaca shares similar concepts with \caesar, as we extended PROV-O to obtain an interoperable description of provenance. However, the provenance information provided by Alpaca is more detailed with respect to the analysis part, which is the main goal of the tool. While \caesar/\ProvBook provides overall descriptions of changes in the source code of \jupyter notebook cells (and the associated results produced by those changes), the details of the functions called inside each cell are not described with the same level of detail as Alpaca. Moreover, although \caesar supports the capture and interoperable serialization of metadata throughout the experiment, Alpaca structures metadata for data objects throughout the code execution during the analysis (e.g., the annotations and attributes of \neo objects), which provides a more fine-grained description of the data evolution (e.g., the removal of the two channels from the data from monkey L in the use case example). In the end, \caesar is a useful tool to capture overall aspects of provenance during the execution of an analysis in the context of an electrophysiology experiment. However, the additional level of detail provided by Alpaca is complementary and could be used to provide additional levels to the provenance, while retaining interoperability.

The \fairworkflows library aims to make workflows implemented within \jupyter notebooks more compliant with the FAIR principles \citep{Richardson21_1}. The library uses decorators to add semantic information to the \python code. After their execution, \fairworkflows constructs RDF graphs describing the workflows using P-Plan \citep{Garijo12} and other ontologies defined by the user in the annotations \citep{Celebi20_e281}. This is linked to the provenance information that is captured during the execution and structured using PROV-O and can be published in the form of nanopublications \citep{Kuhn16_e78}. The use of decorators to instrument the functions is similar to Alpaca, and the decorators of \fairworkflows might be used within scripts such as \scriptfile. However, while Alpaca makes a distinction between inputs, outputs and parameters (from the arguments that a \python function can take and its return values), \fairworkflows makes a direct mapping of arguments as inputs and function returns as outputs. Therefore, the semantic model for provenance in Alpaca emphasizes the identification of the parameters relevant to control the execution of particular functions. For example, in the computation of the PSD using \elephantWelch, \fairworkflows would consider the 2~Hz value an input to the function, when Alpaca records it as the special property \alpacaHasParameter. This is particularly relevant when querying the information using SPARQL, for instance. Moreover, Alpaca also captures and describes detailed information about the objects, which we showed to be relevant for the correct interpretation of the results. However, the extra information from the semantic annotations in \fairworkflows could be combined with Alpaca to provide more descriptive provenance and published using the nanopublication engine.

Computational models are frequently used together with electrophysiology experiments to understand brain function and dynamics. Several state-of-the art simulation engines (e.g., NEST, \cite{Gewaltig_07_11204}; NEURON, \cite{Hines97_1179}; Brian, \cite{Goodman08_5}) are available, and many are implemented in \python or provide high-level \python interfaces where neuronal models with different complexities and biological details can be easily constructed using \python scripts (e.g., by using an interface such as PyNN; \cite{Davison09}). In this context, Alpaca might be useful to track the sequence of functions and respective parameters used to instantiate the models in the simulator and run the simulations. This could be used as a complement to tools such as \sumatra \citep{Davison14_57}, that functions as an electronic lab notebook for simulations, capturing coarse level provenance when executing simulation scripts. Another example is for a tool such as \bennch \citep{Albers22_837549}, that implements a modular workflow for performance benchmarking of neuronal network simulations and could profit from a more fine-grained capture of details in the model and configuration step. Therefore, there is the possibility of also using Alpaca outside of experimental scenarios.

A useful tool for electrophysiology data analysis pipelines is a WMS such as \snakemake \citep{Koster12_2520}. A particularity of \snakemake as a WMS is that it orchestrates the execution of different steps that can take the form of custom \python scripts, instead of modular and specific workflow elements such as the ones provided by a WMS such as LONI Pipeline \citep{MacKenzie-Graham08_208}. This is attractive when working with electrophysiology data as different aspects of the analysis process (as mentioned in Section 1) can be considered yet providing modular and reusable elements \citep{Gutzen22_arxiv}. The \snakemake WMS is based on binding input and output files as dependencies to each script executed in sequence. Therefore, one could envision a scenario where a script such as \scriptfile would have all parameters passed by command line and the execution was controlled by \snakemake. In this scenario, \snakemake would describe the \nix files and the \plotfile output as inputs and output of \scriptfile, respectively, together with the description of the command line parameters. However, this would still rely on the correct mapping of all command line parameters to the actual \python functions (such as the filter cutoff in \elephantButter or frequency resolution in \elephantWelch). Any parameters potentially hard coded directly into the function calls would not be captured and would result in a wrong or incomplete description of provenance. In contrast, all function-level parameters are tracked automatically with Alpaca. Finally, the provenance description of a \snakemake execution in the form of directed acyclic graphs is currently stored in a non-interoperable format. Therefore, it is likely that Alpaca can be a complementary solution to use with \snakemake in more complex analysis scenarios, such as the ones that requires multiple scripts for modularity. However, the provenance description would be enhanced: while the provenance at the file/script level would be provided by \snakemake, the additional metadata file produced by Alpaca would provide a more fine-grained level of detail regarding each step of the workflow, while adding interoperability. 

\subsection{Limitations}

The initial implementation of Alpaca described in this manuscript has some limitations with respect to the scope of the captured provenance. Here, we describe these and suggest remedies. 

First, Alpaca is not capturing and saving information regarding the execution environment such as \python interpreter information, packages installed, operating system, and hardware details. However, there are existing tools that can be used for that purpose and that could be used to run a script instrumented with Alpaca (e.g., \sumatra; \cite{Davison14_57}). Moreover, Alpaca could be integrated with such tools to use the information provided by them in the saved provenance records. In the end, we focused on adding granularity instead of reimplementing functionality of existing tools, as this information is more relevant for understanding and sharing the electrophysiology analysis result.

Second, the Alpaca ontology is currently not structured to allow the description of the execution environment. It could be further expanded to include any information regarding the environment, as one could envision a revised Alpaca provenance model and ontology with a PROV Agent subclass that would be related to \alpacaScriptAgent, and whose properties would describe the relevant aspects of the environment. Moreover, the description could be further improved by integration with other ontologies developed specifically for the detailed description of experimental workflows, such as P-Plan \citep{Garijo12} and REPRODUCE-ME \citep{Samuel22_1}. Therefore, although not present in this initial implementation, the approach adopted allows easy expansion and integration of additional features.

Third, some steps are visible from the data flow perspective but they are not fully descriptive and understandable at this point. One example are user-defined functions, such as \scriptPlotLfpPsd in \scriptfile. As a plotting function, the user might be interested in knowing additional details on how the inputs (i.e., the \matplotlib \matplotlibAxes object and the arrays with the data) were handled. The current implementation tracks code in a single scope, and therefore the execution of a function such as \scriptPlotLfpPsd is treated as a "black box". It would be interesting to also capture the execution of some functions with an even finer description of the operations inside those functions. This could be achieved by expanding the functionality to automatically include functions in levels lower than the primary capture scope. However, even in the current implementation of Alpaca, although such fine descriptions from inside of \scriptPlotLfpPsd are not available, the provenance stored in the generated metadata file already points to where the function was implemented. In this way, the user can focus on inspecting the implementation of the function \scriptPlotLfpPsd and does not have to check the full source code.

Finally, only a generic visualization graph is currently provided in Alpaca. Although we took the approach to leverage the advantage of open-source graph visualization tools such as \gephi, the visualization of the captured provenance is not optimized (e.g., showing only parameters of the selected function or object). It is important to note that this can be incorporated as additional feature in Alpaca without any changes to the captured information or serialization as RDF, by using existing graph visualization frameworks such as \textit{Pyvis} to build a customized visualization environment based on the information in the RDF graphs. Finally, there are existing tools that specifically deal with the visualization of provenance graphs. One example is \avocado \citep{Stitz16_481}, implemented to be an interactive provenance graph visualization tool that exploits the topological structure of the graph to provide a visual aggregation. Although Alpaca provides basic aggregation using functionality adapted from \networkx, we could also leverage a tool like \avocado to provide visualization functionality more tailored to the features of a provenance graph, such as hierarchical structure (e.g., all the steps in a single trial-processing loop grouped in a single node) and temporal evolution (isolating the visualization of the analyses performed in the first or the second dataset). However, the technical challenges of such integration are unknown at this point. 

\subsection{Future directions}

Several improvements are planned for Alpaca in the future. First, we plan to expand the toolbox to also capture provenance for analyses implemented using \jupyter notebooks. Not only is \jupyter extensively used for exploratory data analysis, but also the repeated execution of code cells and subsequent substitution of data objects in memory requires detailed provenance tracking for reliable description of any analysis result produced by a notebook. Second, the provenance records lack semantic information that are relevant for understanding electrophysiology data and metadata. Therefore, a further improvement is to allow the inclusion of classes and vocabularies defined in domain-specific ontologies in the provenance records, which will bring further improvements to the FAIRness of electrophysiology analysis results. As discussed above, the functionality will also be improved to capture information about the execution environment, together with information from version control systems such as \git, to provide more detailed information about the source code that originated the analysis result. Finally, we aim to further improve the interaction and analysis of the captured provenance by developing a custom visualization and search interface based on the serialized RDF graphs. 

\subsection{Conclusions}

We implemented Alpaca, a toolbox for lightweight provenance capture during the execution of \python scripts used for the analysis of electrophysiology data. Alpaca captures more detailed information about the analysis processes, including not only the lineage of the data but also embedded metadata relevant for the description of data objects during the processing pipeline. In the end, this makes the electrophysiology analysis result artefacts more compliant to the FAIR principles. This may improve research reproducibility and the trust in the results, especially in collaborative environments. Therefore, Alpaca may be a valuable tool to facilitate sharing electrophysiology data analysis results. 

%
%

\section*{Acknowledgements}
We thank Andrew Davison and Richard Gerkin for helpful discussions during the implementation of Alpaca. We also thank Oliver Kloß for testing Alpaca. We thank Angela Fischer for helping to construct cartoon images in Figure 1. Human figures were custom-designed based on publicly and freely available images on Unsplash and Pixabay.

\section*{Funding}
\printfunding

\section*{Author contributions}
Conceptualization: C.K., D.U., S.G., S.D., M.D.; Methodology: C.K., S.D., M.D.; Software: C.K.; Writing - original draft: C.K., S.G., M.D.; Writing - review \& editing: C.K., D.U., S.G., S.D., M.D.; Supervision: M.D., S.G., S.D.; Funding acquisition: M.D., S.G.
\\

%
%

\printbibliography

\end{document}


\onecolumn

\maketitle

\section{Changes to the code necessary to track provenance using Alpaca}

The script \scriptfile{} was modified to:
\begin{itemize}
	\item apply the decorator to functions imported from other modules or to methods inside classes defined in other modules;
	\item apply the decorator to user-defined functions in the script (i.e., using the Python \textit{def} keyword);
	\item activate provenance tracking at the beginning of the \textit{main} function;
	\item serialize the provenance trace as a Turtle file at the end of the \textit{main} function.
\end{itemize}

This output is produced by running \textit{git diff --no-index ./code/no\_provenance/psd\_by\_trial\_type.py ./code/provenance/psd\_by\_trial\_type.py}, which are the versions of \scriptfile{} without and with Alpaca provenance tracking, respectively. These scripts are available at the accompanying archive of this paper.

\bigskip{}

\ttfamily{}

\begin{Verbatim}[commandchars=\\\{\}]

\textbf{diff --git a/./code/no_provenance/psd_by_trial_type.py b/./code/provenance/psd_by_trial_type.py}
\textbf{index 6bc7f5d..f45e70e 100644}
\textbf{--- a/./code/no_provenance/psd_by_trial_type.py}
\textbf{+++ b/./code/provenance/psd_by_trial_type.py}
@@ -17,12 +17,37 @@ import matplotlib.pyplot as plt
 from numpy import mean
 from scipy.stats import sem
 
\textcolor{ForestGreen}{+from alpaca import Provenance, activate, save_provenance, alpaca_setting}
\textcolor{ForestGreen}{+from alpaca.utils.files import get_file_name}
\textcolor{ForestGreen}{+}
\textcolor{ForestGreen}{+}
\textcolor{ForestGreen}{+# Apply the decorator to the functions used}
\textcolor{ForestGreen}{+}
\textcolor{ForestGreen}{+butter = Provenance(inputs=['signal'])(butter)}
\textcolor{ForestGreen}{+}
\textcolor{ForestGreen}{+welch_psd = Provenance(inputs=['signal'])(welch_psd)}
\textcolor{ForestGreen}{+}
\textcolor{ForestGreen}{+add_epoch = Provenance(inputs=['segment', 'event1', 'event2'])(add_epoch)}
\textcolor{ForestGreen}{+}
\textcolor{ForestGreen}{+get_events = Provenance(inputs=['container'],}
\textcolor{ForestGreen}{+                        container_output=True)(get_events)}
\textcolor{ForestGreen}{+}
\textcolor{ForestGreen}{+cut_segment_by_epoch = Provenance(inputs=['seg', 'epoch'],}
\textcolor{ForestGreen}{+                                  container_output=True)(cut_segment_by_epoch)}
\textcolor{ForestGreen}{+}
\textcolor{ForestGreen}{+mean = Provenance(inputs=['a'])(mean)}
\textcolor{ForestGreen}{+}
\textcolor{ForestGreen}{+sem = Provenance(inputs=['a'])(sem)}
\textcolor{ForestGreen}{+}
\textcolor{ForestGreen}{+neo.AnalogSignal.downsample = Provenance(inputs=['self'])(neo.AnalogSignal.downsample)}
\textcolor{ForestGreen}{+}
 
 # Setup logging
 logging.basicConfig(level=logging.INFO,
                     format="[%(asctime)s] %(name)s %(levelname)s: %(message)s")
 
 
\textcolor{ForestGreen}{+@Provenance(inputs=[], file_input=['file_name'])}
 def load_data(file_name):
     """
     Reads all blocks in the NIX data file `file_name`.
@@ -33,6 +58,7 @@ def load_data(file_name):
     return block
 
 
\textcolor{ForestGreen}{+@Provenance(inputs=[])}
 def create_main_plot_objects(n_subjects, title):
     """
     Creates the plotting grid and figure/axes, with proper distribution and
@@ -62,6 +88,7 @@ def create_main_plot_objects(n_subjects, title):
     return fig, axes
 
 
\textcolor{ForestGreen}{+@Provenance(inputs=['signal'])}
 def select_channels(signal, skip_channels):
     """
     Selects all the channels from the AnalogSignal `signal` that are not
@@ -75,6 +102,7 @@ def select_channels(signal, skip_channels):
     return signal[:, mask]
 
 
\textcolor{ForestGreen}{+@Provenance(inputs=['axes', 'freqs', 'psds', 'sem'])}
 def plot_lfp_psd(axes, freqs, psds, sem, label, sem_multiplier=1.96,
                  freq_range=None, **kwargs):
     """
@@ -99,6 +127,7 @@ def plot_lfp_psd(axes, freqs, psds, sem, label, sem_multiplier=1.96,
     return axes
 
 
\textcolor{ForestGreen}{+@Provenance(inputs=['axes', 'title'])}
 def set_title(axes, title):
     """
     Set the title of the plot in `axes`.
@@ -108,6 +137,7 @@ def set_title(axes, title):
     axes.set_title(title)
 
 
\textcolor{ForestGreen}{+@Provenance(inputs=['fig'], file_output=['file_name'])}
 def save_plot(fig, file_name, **kwargs):
     """
     Save the plot in `fig` to the file `file_name`.
@@ -115,6 +145,7 @@ def save_plot(fig, file_name, **kwargs):
     fig.savefig(file_name, **kwargs)
 
 
\textcolor{ForestGreen}{+@Provenance(inputs=['arrays'])}
 def vstack_quantities(*arrays):
     """
     Performs stacking of Quantity arrays, as ordinary `np.vstack` removes
@@ -144,6 +175,14 @@ def vstack_quantities(*arrays):
 
 
 def main(session_files, output_dir, skip_channels):
\textcolor{ForestGreen}{+}
\textcolor{ForestGreen}{+    # Use builtin hash for matplotlib objects}
\textcolor{ForestGreen}{+    alpaca_setting('use_builtin_hash_for_module', ['matplotlib'])}
\textcolor{ForestGreen}{+    alpaca_setting('authority', "fz-juelich.de")}
\textcolor{ForestGreen}{+}
\textcolor{ForestGreen}{+    # Activate provenance tracking}
\textcolor{ForestGreen}{+    activate()}
\textcolor{ForestGreen}{+}
     logging.info(f"Processing files: {','.join(session_files)}")
 
     # Labels of the different trial types in the data, and the colors
@@ -236,6 +275,11 @@ def main(session_files, output_dir, skip_channels):
     logging.info(f"Saving output to {out_file}")
     save_plot(fig, out_file, format='png', dpi=300)
 
\textcolor{ForestGreen}{+    # Save provenance information as Turtle file}
\textcolor{ForestGreen}{+    prov_file_format = "ttl"}
\textcolor{ForestGreen}{+    prov_file = get_file_name(out_file, extension=prov_file_format)}
\textcolor{ForestGreen}{+    save_provenance(prov_file, file_format=prov_file_format)}
\textcolor{ForestGreen}{+}
 
 if __name__ == "__main__":
     parser = argparse.ArgumentParser()

\end{Verbatim}